\documentclass[traditabstract]{aa}  

\usepackage{graphicx}
\usepackage{tabularx,booktabs}
\usepackage{grffile}
\usepackage[varg]{txfonts}
\usepackage{subfigure}
\usepackage{amssymb}
\usepackage{cancel}
\usepackage{bm}
\usepackage[draft]{hyperref}
\usepackage{color}
\usepackage{footnote}
\usepackage{eqnarray}
\usepackage{soul}
\newcolumntype{Y}{>{\centering\arraybackslash}X}
\begin{document}


   \title{Annular substructures in the transition disks \\ around LkCa~15 and J1610}
   
\titlerunning{Annular substructures in transition disks}

   \author{S.~Facchini\inst{1}   
           \and
           M.~Benisty\inst{2,3,4}
           \and
           J.~Bae\inst{5}
           \and
           R.~Loomis\inst{6}
           \and
           L.~Perez\inst{2}
           \and
           M.~Ansdell\inst{7}
           \and
           S.~Mayama\inst{8}
           \and
           P.~Pinilla\inst{9}
           \and
           R.~Teague\inst{10}
           \and
           A.~Isella\inst{11}
           \and
           A.~Mann\inst{12}
          }


   \institute{European Southern Observatory, Karl-Schwarzschild-Str. 2, 85748 Garching, Germany
   \\
   \email{stefano.facchini@eso.org} 
   \and
   Departamento de Astronom\'ia, Universidad de Chile, Camino El Observatorio 1515, Las Condes, Santiago, Chile
   \and
   Unidad Mixta Internacional Franco-Chilena de Astronom\'ia, CNRS, UMI 3386
   \and
   Univ. Grenoble Alpes, CNRS, IPAG, 38000 Grenoble, France.
   \and
   Earth and Planets Laboratory, Carnegie Institution for Science, 5241 Broad Branch Road NW, Washington, DC 20015,
USA
   \and
   NRAO, 520 Edgemont Rd, Charlottesville, VA 22903, USA
   \and
   Flatiron Institute, Simons Foundation, 162 Fifth Ave, New York, NY 10010, USA
   \and
   The Graduate University for Advanced Studies, SOKENDAI, Shonan Village, Hayama, Miura, Kanagawa 240-0193, Japan
   \and
   Max Planck Institute for Astronomy, K\"onigstuhl 17, 69117, Heidelberg, Germany
   \and
   Center for Astrophysics | Harvard \& Smithsonian, 60 Garden Street, Cambridge, MA 02138, USA
   \and
   Department of Physics and Astronomy, Rice University, 6100 Main Street, MS-108, Houston, TX 77005, USA
   \and
   Department of Physics and Astronomy, University of North Carolina at Chapel Hill, Chapel Hill, NC 27599, USA
   }

   \date{Received; accepted}

 
  \abstract
   {We present high resolution millimeter continuum ALMA observations of the disks around the T Tauri stars LkCa~15 and 2MASS~J16100501-2132318 (hereafter, J1610). These transition disks host dust-depleted inner regions, possibly carved by massive planets, and are of prime interest to study the imprints of planet-disk interactions. While at moderate angular resolution they appear as a broad ring surrounding a cavity, the continuum emission resolves into multiple rings at a resolution of $\sim$60$\times$40\,mas ($\sim7.5\,$au for LkCa~15, $\sim6\,$au for J1610) and $\sim7\,\mu$Jy\,beam$^{-1}$ rms at $1.3\,$mm. In addition to a broad extended component, LkCa~15 and J1610 host 3 and 2 narrow rings, respectively, with two bright rings in LkCa~15 being radially resolved. LkCa~15 possibly hosts another faint ring close to the outer edge of the mm emission. The rings look marginally optically thick, with peak optical depths of $\sim0.5$ (neglecting scattering), in agreement with high angular resolution observations of full disks. We perform hydrodynamical simulations with an embedded, sub-Jovian-mass planet and show that the observed multi-ringed substructure can be qualitatively explained as the outcome of the planet-disk interaction. We note however that the choice of the disk cooling timescale alone can significantly impact the resulting gas and dust distributions around the planet, leading to different numbers of rings and gaps and different spacings between them. We propose that the massive outer disk regions of transition disks are favorable places for planetesimals and possibly second generation planet formation of objects with a lower mass than the planets carving the inner cavity (typically few $M_{\rm Jup}$), and that the annular substructures observed in LkCa~15 and J1610 may be indicative of planetary core formation within dust-rich pressure traps. Current observations are compatible with other mechanisms being at the origin of the observed substructures, in particular with narrow rings generated (or facilitated) at the edge of the CO and N$_2$ snowlines.
   }

   \keywords{accretion, accretion disks -- planetary systems: protoplanetary disks -- stars: individual: LkCa15 -- stars: individual: J1610 -- submillimeter: planetary systems 
               }

   \maketitle
%
\section{Introduction}
\label{sec:intro}

High angular resolution observations of protoplanetary disks show that substructure in their dust emission is ubiquitous. This is the case in both (sub-)millimeter (mm) emission, largely tracing dust grains with sizes $>100\,\mu$m \citep[e.g.,][]{andrews2020}, and in scattered light observations at optical and near infrared (NIR) wavelengths, probing sub-$\mu$m sized dust grains \citep[e.g.,][]{garufi2018}. The only disks that so far do not show azimuthal or radial substructure in their dust distribution are very compact or faint objects, where the finite resolution of the instruments is able to image these disks with just a few (or less) resolution elements across their diameter \citep[e.g.,][]{facchini2019}. Before the advent of the high resolution Atacama Large Millimeter/submillimeter Array (ALMA) and extreme adaptive optics systems on ground based telescopes, photometric surveys had already identified a class of objects with some level of substructure, the so-called transition disks, evidenced from a lack of near/mid-IR emission in their spectral energy distribution that indicates a dust-depleted inner region. This was confirmed with sub-mm images that clearly showed the presence of cavities with angular sizes larger than the resolution elements available at the time \citep[0.3\arcsec, $\sim$40-75\,au;][]{brown2009,andrews2011} surrounded by a bright and wide ring. Their morphology has been explained as the outcome of planet-disk interactions, with massive companions carving a cavity and generating a pressure maximum outside their orbital radius where mm grains drift towards and accumulate \citep[e.g.,][]{rice2006}. In some cases, the radial extent of the cavities is too large to be explained by a single planet orbiting on a circular orbit and the presence of multiple giant planets was suggested \citep{dodson2011, bae2019}. Other physical mechanisms have also been invoked to explain the lack of dust particles in the inner regions of transition disks, such as internal photoevaporation \citep[e.g.][]{alexander_06,gorti2009,owen_10} and dead zones \citep[e.g.][]{regaly2012,flock2015}.

High angular resolution observations of these objects at mm wavelengths are crucial to understand the origin of these cavities. The radial and azimuthal brightness distributions of the ring itself can be analyzed as probes of planet-disk interactions, with e.g., asymmetries tracing vortices due to the Rossby Wave instability. Independently of the mechanism clearing the central cavity, it is clear that dust trapping must be occurring in the ring, since mm dust grains would otherwise naturally drift towards the central star. In these rings, the dust-to-gas ratio is expected to be close to unity due to the efficient trapping at the pressure maximum. In such physical conditions, planetesimal formation is thought to occur efficienty via the streaming-instability \citep[e.g.,][]{youdin2005,johansen2007}. Subsequent triggering of core-formation via pebble accretion may occur \citep[e.g.,][]{ormel2010,bitsch2015}, even though theoretical models suggest that the growth timescales could be too long for this process to be efficient at large distances from the star within the disk lifetime \citep[e.g.][]{morbidelli2020}. These rings are likely among the best environments to look for signatures of on-going formation of planetesimals and possibly planetary cores.

\begin{table*}
\centering
\caption{Log of ALMA observations used in this paper.}
\begin{tabularx}{\textwidth}{lc c*{6}{Y}}
\toprule
Object & Date	&  Antennas & Min. baseline	& Max. baseline	 & Time on source & Bandpass calibrator & Phase calibrator & Flux calibrator	\\
\midrule
LkCa~15 	& 18.11.2018	& 45 & 15\,m & 1397\,m & 31\,min & J0510+1800 & J0426+2327 & J0510+1800 \\
LkCa~15 	& 13.07.2019	& 40 & 111\,m & 12644\,m & 37\,min & J0510+1800 & J0431+2037 & J0510+1800 \\
LkCa~15 	& 19.07.2019	& 43 & 96\,m & 8547\,m & 37\,min & J0510+1800 & J0431+2037 & J0510+1800 \\
\midrule
J1610 	& 03.09.2019	& 46 & 38\,m & 3143\,m & 19\,min & J1517-2422 & J1551-1755 & J1517-2422 \\
J1610 	& 17.07.2019	& 43 & 92\,m & 8547\,m & 43\,min & J1427-4206 & J1551-1755 & J1427-4206 \\
J1610 	& 18.07.2019	& 46 & 92\,m & 8547\,m & 19\,min & J1427-4206 & J1551-1755 & J1427-4206 \\
\midrule
\bottomrule
\end{tabularx}
\label{tab:observations}
\end{table*}

Some transition disks have been imaged at high angular resolution at mm wavelengths  ($\sim20-70\,$mas), showing that the wide rings observed at intermediate resolution are actually composed of smaller-scale features. They show complex morphologies and finer levels of substructure in both radial and azimuthal directions. Multiple rings, spirals, arcs and eccentric features have been detected, showing high level of complexity  \citep[e.g.,][]{dong2018,andrews2018,casassus2019,rosotti2019,perez2019}. Interestingly, in the handful of disks that have been observed at similarly high resolution in both scattered light and mm thermal emission, there is no strict correspondence between the morphological features \citep[e.g.,][]{cazzoletti2018}, indicating that the physical processes responsible for these complex morphologies may have different imprints depending on the disk tracer. 

In this paper, we present new ALMA Band 6 220 GHz continuum observations of LkCa~15 and 2MASS J16100501-2132318 (EM*StHA 123, EPIC~204630363, hereafter J1610) at unprecedented angular resolution ($\sim40\times60\,$mas, i.e. $\sim7.5$\,au and $6\,$au in radius for LkCa~15 and J1610, respectively). Both objects are classical T Tauri stars hosting transition disks with resolved cavities \citep{pietu2006,ansdell2020}. Interestingly they are also classified as dippers, based on the short term variability seen in their optical light curves \citep{ansdell2016,rodriguez2017,alencar2018}. This photometric variability is interpreted as due to a warp in a misaligned inner disk, possibly due to an inclined magnetic field or a companion on an inclined orbit.

LkCa~15 is a well-studied object, which garnered a lot of attention due to the claims of giant planets in the disk inner cavity \citep{kraus_12,sallum2015}. Recent results show that the emission is however likely due to disk signal \citep{thalmann2016,mendigutia2018,currie2019}. Located at a distance of $158.9\pm1.2\,$pc \citep{gaia}, LkCa~15 is a $1.25\pm0.10\,M_\odot$ star, with an age of $\sim5\,$Myr and accretion rate of $10^{-9.2\pm0.3}\,M_{\odot}/$yr and a stellar bolometric luminosity of $1.05^{+0.27}_{-0.21}\,L_{\odot}$ \citep{donati2019}. Note that previous works used a stellar luminosity of $0.74\,L_\odot$ derived with the pre-Gaia distance. Dynamical estimates of the central star matches well to the one estimated from the pre-main sequence track \citep{qi2019}. The disk is very bright in the mm \citep[380\,mJy at $870\,\mu$m; ][]{andrews2011}, with a dust disk mass of $\sim165\,M_\oplus$ in the optically thin assumption \citep{isella2012}, and a gaseous disk traced by CO out to $\sim900\,$au \citep[e.g.,][]{jin2019}. While the disk exhibits a clear dust-depleted cavity at mm wavelengths, of $\sim50\,$au in radius \citep[e.g.,][]{pietu2006,isella2012,pinilla2018}, scattered light observations indicate the presence of an inner disk extending up to $\sim$30\,au within the mm cavity \citep[e.g.,][]{thalmann2016,oh2016}. Such a spatial segregation in dust sizes is a natural outcome of planet-disk interactions \citep{pinilla_12} and supports the presence of a massive planet orbiting at $\sim$40\,au.

J1610 is a less studied object. \citet{rizzuto2015} spectroscopically identified it as a low-mass member of the Upper Sco association. It is located at a distance of $144.7\pm2.7\,$pc \citep{gaia} and the star has a K7.5 spectral type, with evidence of accretion from H$\alpha$ emission \citep{ansdell2016}. Using the new Gaia distance, we fit the SNIFS stellar spectrum and broadband photometric spectral energy distribution (SED) from \citet{ansdell2016} and obtain a stellar luminosity of $0.46\pm0.03\,L_{\odot}$, with $T_{\rm eff}=3950\pm80\,$K (more details in Appendix \ref{sec:app_sed}). By using the \citet{siess2000} pre-main sequence evolutionary tracks as done by \citet{barenfeld2016} for a large sample of Class II objects in Upper-Scorpius, we obtain a stellar mass of $0.67\pm0.11\,M_\odot$, where we assumed a metallicity $Z=0.02$ and no convective overshoot.  The disk has been detected with the Sub-millimeter Array (SMA) with an integrated flux density of $28.0\pm1.5\,$mJy at 1.3\,mm \citep{ansdell2016}, with estimated dust mass of $15\,M_\oplus$ in the optically thin and isothermal assumptions, and has been shown to have a cavity of $\sim20\,$au in radius from previous moderate resolution ALMA observations \citep{ansdell2020}. 

The paper is organized as follows. In Sec.~\ref{sec:data} the observations, data calibration and imaging parameters are presented. Sec.~\ref{sec:results} summarizes the data analysis, and Sec.~\ref{sec:hydro} compares the data with hydrodynamical simulations of planet-disk interaction. In Sec.~\ref{sec:discussion} we discuss our results and draw our conclusions in Sec.~\ref{sec:conclusions}. 

\begin{figure*}[t]
\begin{center}
\includegraphics[width=0.49\textwidth]{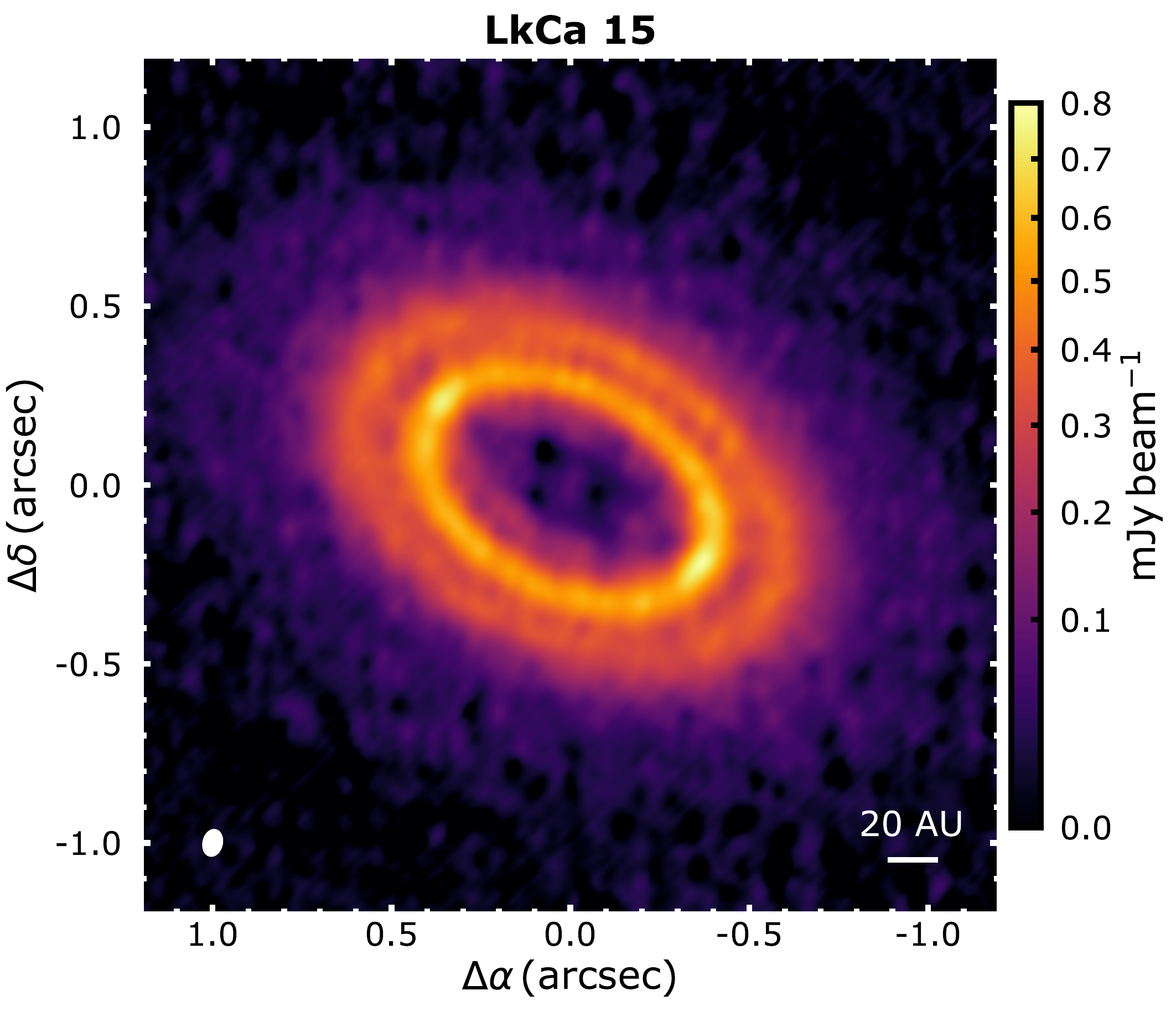}
\includegraphics[width=0.49\textwidth]{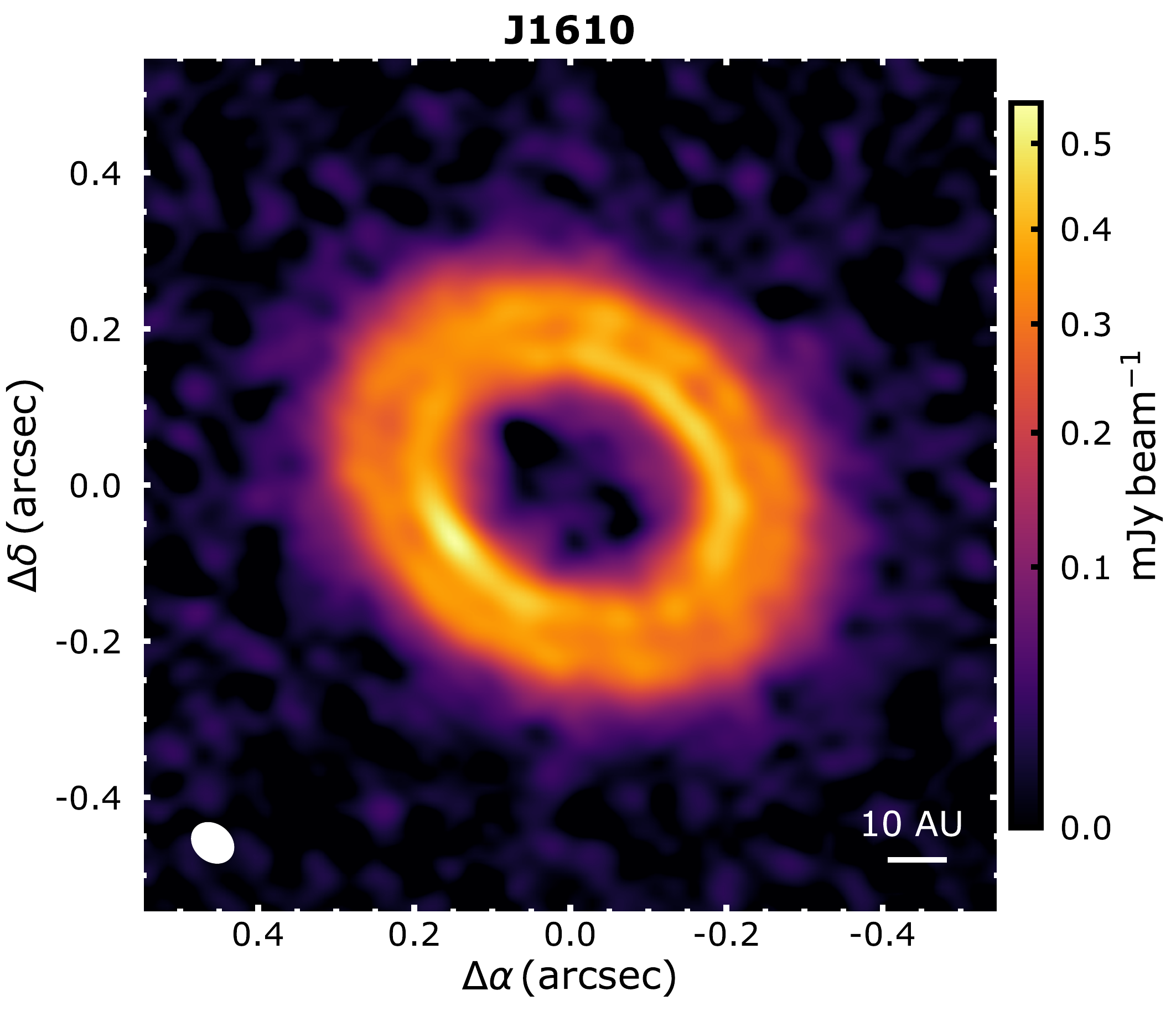}
\end{center}
\caption{Intensity maps of LkCa~15 (left panel) and J1610 (right panel). Note the different angular scales of the two panels. The color scale has been stretched to highlight the fainter regions of the disks.}
\label{fig:maps}
\end{figure*}

\section{ALMA observations and data reduction}
\label{sec:data}

LkCa~15 and J1610 were observed in Band 6 as part of the ALMA Program \#2018.1.01255.S. The observations were carried out with different configurations in order to provide good uv-coverage at different spatial frequencies, using nominal C43-5, C43-8 and C43-9 configurations for LkCa~15, and C43-6 and C43-8 for J1610. Details of the observations are reported in Table~\ref{tab:observations}. The maximum recoverable scale for LkCa~15 and J1610 is $3.7\arcsec$ and $1.8\arcsec$, respectively. The spectral setup had 4 spectral windows, three dedicated for continuum observations in TDM mode, and one dedicated for high spectral resolution observations of the $^{12}$CO line. In this paper we focus on the three continuum spectral windows only, which have central rest frequency of 214, 216.2 and 229\,GHz.  

The data were calibrated using the {\tt CASA} package, version 5.6 \citep{2007ASPC..376..127M}. Self-calibration was performed on all data-sets, leading to a good improvement in the signal-to-noise (S/N) ratio. The visibilities were merged using the {\tt concat} task in {\tt CASA}, and spectrally re-binned on $250\,$MHz channels to avoid bandwidth smearing. 
Images in the sky plane were produced using the {\tt tclean} task, with the {\tt multiscale} cleaning algorithm allowing for point source emission. Elliptical masks were applied, with position angle and inclination derived from $(u,v)$-plane analysis (see Sec.~\ref{sec:galario}), and a semi-major axis of $1.7\arcsec$ and $1\arcsec$ for LkCa~15 and J1610, respectively. The de-convolution was performed down to a cleaning threshold of $1\sigma$, which maximises the flux in the clean model. The residuals in the final images were rescaled by the ratio of the clean beam and dirty beam \citep{JvM1995}, which reduced the rms noise level by $\sim50\%$. This is to correct for the fact that the final image is the sum of the restored clean components (in units of clean beams) and of the residuals (in units of dirty beams); to properly estimate the flux of the residuals, a re-normalization factor equal to the ratio of the two beam areas is applied to them. This method has been successfully applied in Very Large Array (VLA) surveys such as THINGS \citep{walter2008}, and more recently by \citet{Pinte2020} on ALMA data of protoplanetary disks. Different weighting schemes were tested to produce the images. The best compromise between angular resolution and S/N for LkCa~15 is with a Briggs robust weighting  of 0, whereas for J1610 we opted for a Briggs robust weighting of 0.3. The resulting synthesized beam for LkCa~15 is $68\times47\,$mas, with a position angle (P.A.) of $347.4^\circ$. The rms noise level is $\sim6.9\,\mu$Jy\,beam$^{-1}$, as estimated from an annulus centered in the phase center with a $2.5-4\arcsec$ range in radii, and the image has a signal-to-noise (S/N) ratio of $\sim115$ at the peak. For J1610, the synthesized beam is $55\times43\,$mas with a P.A. of $50.3^\circ$, and the rms noise level is $\sim7.2\,\mu$Jy\,beam$^{-1}$. The S/N at the peak is $\sim75$. The recovered flux density within the cleaning mask is $136.4\pm0.1\,$mJy and $30.8\pm0.1\,$mJy for LkCa~15 and J1610, respectively. For both disks, the integrated flux densities are in agreement with previous observations at the same wavelength with different interferometers within the 10\% calibration uncertainties, in particular the IRAM PdBI array and CARMA for LkCa~15 \citep{pietu2006,isella2012} and SMA for J1610 \citep{ansdell2016}. The resulting images are shown in Fig.~\ref{fig:maps}.

\begin{figure*}
\begin{center}
\includegraphics[width=0.33\textwidth]{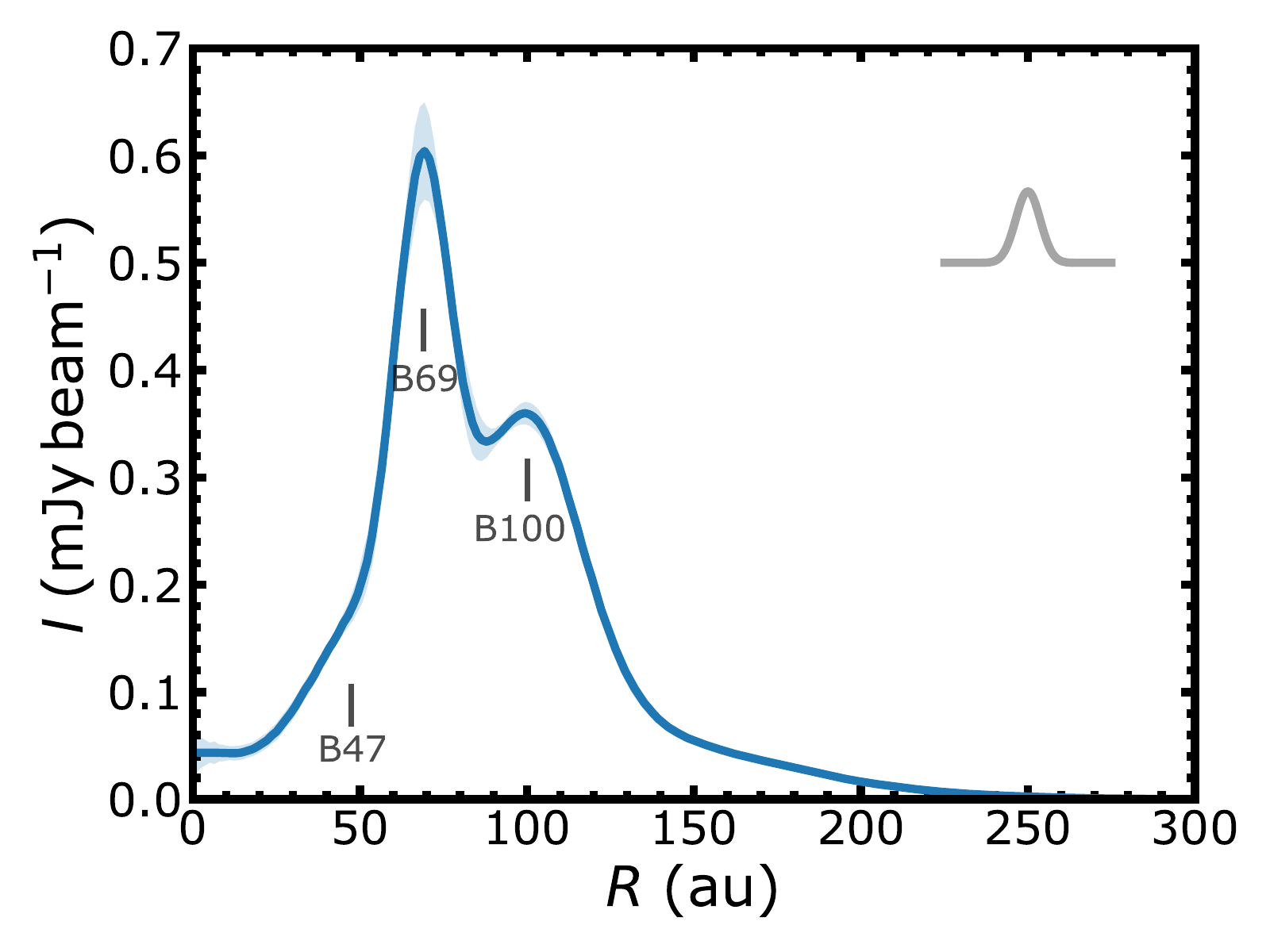} 
\includegraphics[width=0.33\textwidth]{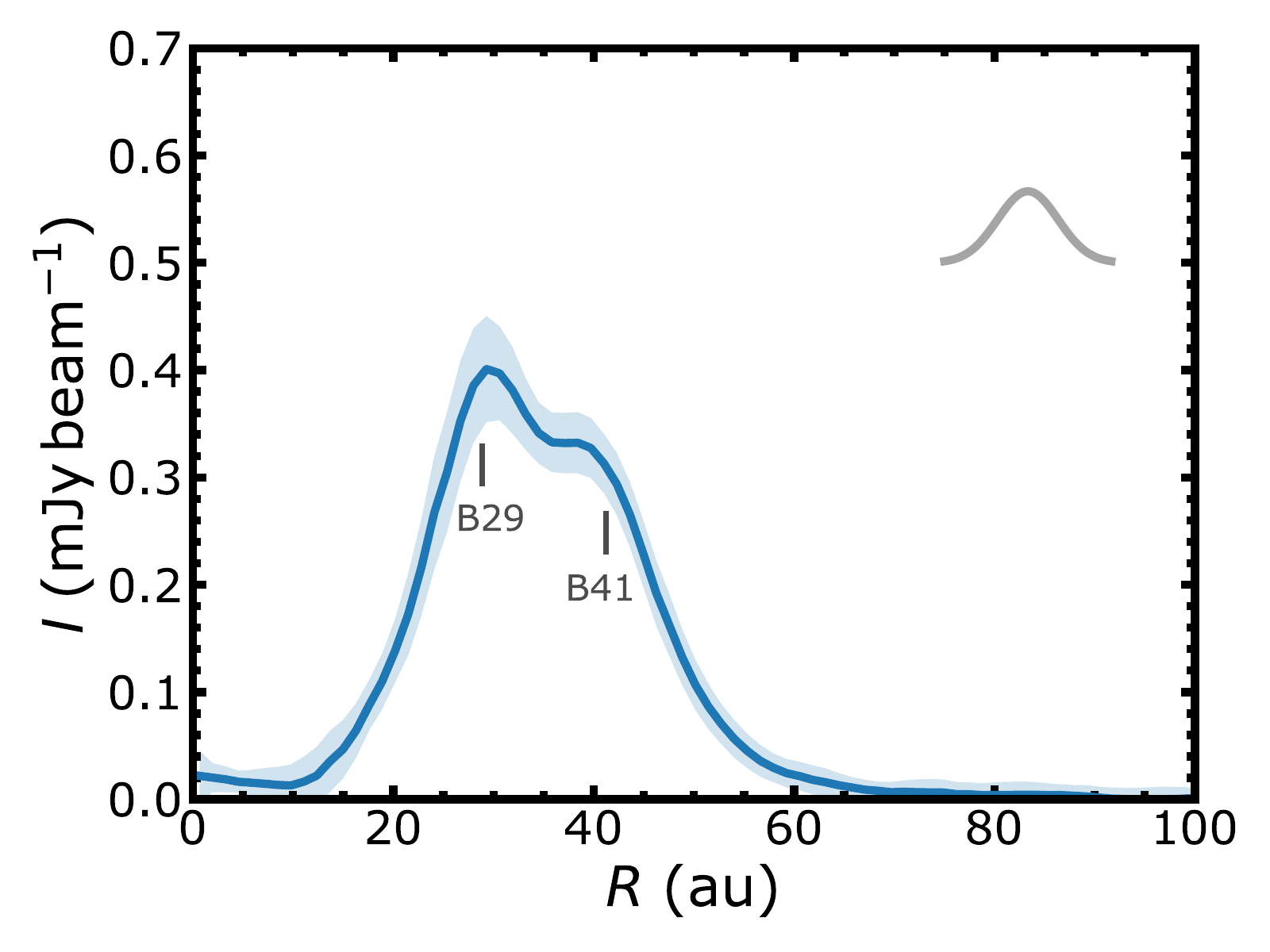}  
\includegraphics[width=0.33\textwidth]{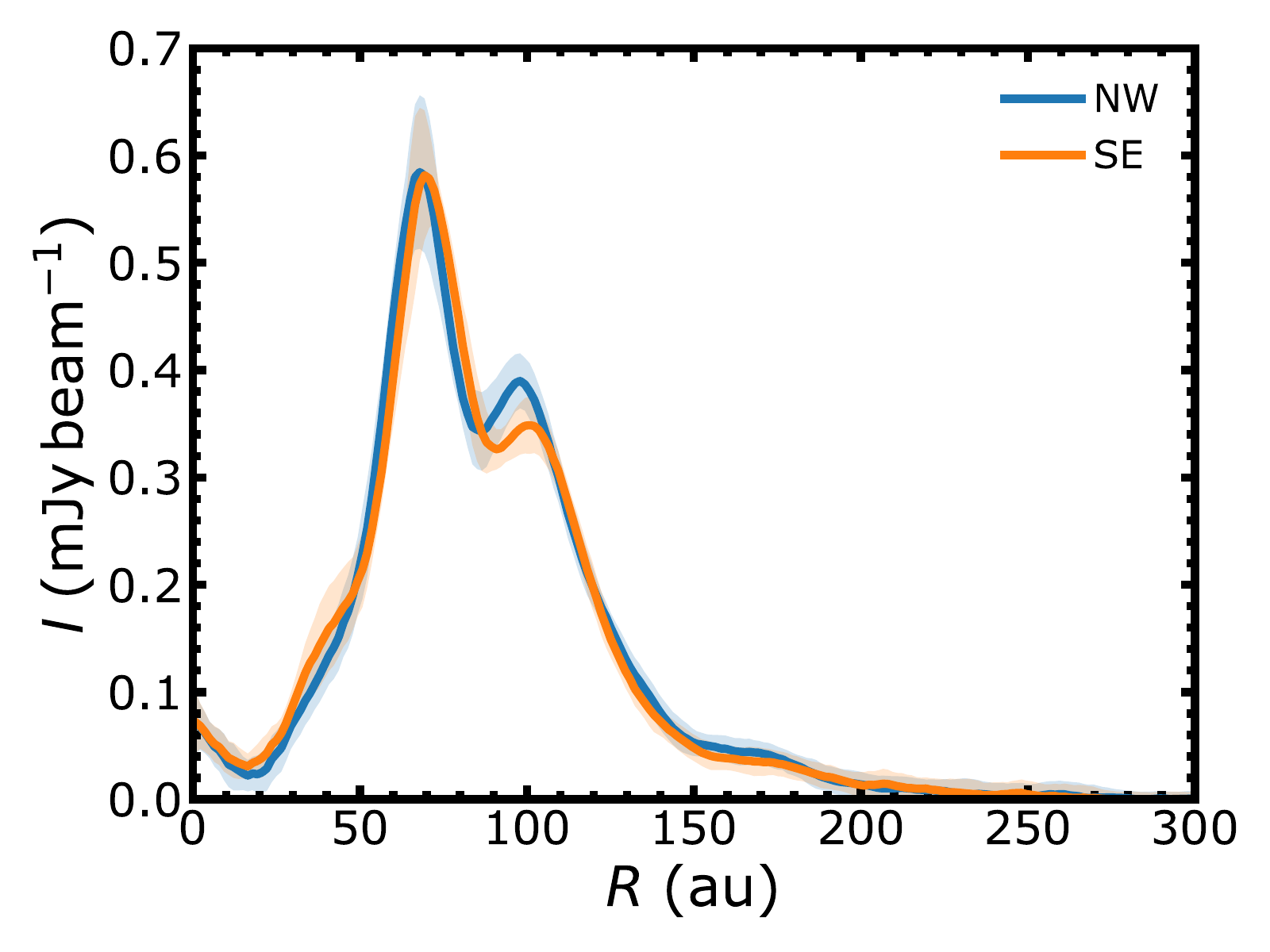}
\end{center}
\caption{Left and middle: deprojected and azimuthally averaged radial intensity profile of the continuum emission of LkCa~15 (left) and J1610 (middle). The ribbon shows the quadratic sum of the standard deviation of the intensity across pixels in each radial bin and the rms of the observations divided by the square root of the number of independent beams sampling the same radial bin in the azimuthal direction. The Gaussian profile in the top right corner of the panels shows the FWHM of the average of major and minor axis of the synthesized beam. Right: azimuthal average of the South-East, North-West sides of LkCa~15, respectively, where the disk has been divided in two sub-regions along the major axis. The inner shoulder is more prominent in the South-East side of the disk.}
\label{fig:rad_prof}
\end{figure*}

\section{Results}
\label{sec:results}
\subsection{Images}
\label{sec:images}
The intensity maps reported in Fig.~\ref{fig:maps} clearly show that the two transition disks have substructure, with the bright mm ring observed in earlier observations separating into distinct multiple rings. Two clear rings are easily observable by looking at the images, and they look remarkably similar in the two disks. One major difference is that the rings observed in J1610 would lay within the cavity of LkCa~15, due to their different angular scales. Compared to other transition disks observed at high angular resolution, these two disks do not show high level of asymmetry, lacking prominent arcs and spirals features  as observed for example in massive disks around Herbig stars as HD135344B \citep{cazzoletti2018} and MWC758 \citep{dong2018}. However, both disks present azimuthal asymmetries along the North-West -- South-East direction, and they are particularly evident in J1610. 

\paragraph{Radial profiles.} In order to analyze the intensity profile, the intensity maps are re-centered, deprojected and azimuthally averaged using the geometrical parameters reported in Tables~\ref{tab:LkCa15}-\ref{tab:J1610} and determined from fitting the visibilities. Details of the geometrical parameters determination can be found in Sec.~\ref{sec:galario}. Radial bins of $9\,$mas are used, and errors are computed as the quadratic sum of the standard deviation of the intensity across pixels in each radial bin and the rms of the observations divided by the square root of the number of independent beams sampling the same radial bin in the azimuthal direction.

The resulting intensity profiles are reported in Fig.~\ref{fig:rad_prof} and show radial substructure, with two narrow rings emerging in both LkCa~15 and J1610, and an underlying extended broad component. A figure with the same profiles in logarithmic scale  highlights the faint extended emission (Fig.~\ref{fig:rad_prof_log}). The intensity peak of the outer ring is always fainter than the inner ring peak, which can be explained by a drop in dust temperature and/or optical depth. On average, the outer ring is brighter in the NW side of the disk by $\sim10\%$ (Fig.~\ref{fig:rad_prof}, right) with the NW side being the close side of the disk to the observer \citep{thalmann2015}. This same effect seems to be even more pronounced in lower resolution band 9 data \citep{vandermarel2015}, and may be indicative of azimuthally asymmetric illumination of the disk due to shadowing \citep{thalmann2016}.

In LkCa~15, the inner ring shows a `shoulder' extending inwards in the cavity, similarly to other transition disks imaged at high angular resolution \citep[e.g.,][]{perez2019,huang2020}. The shoulder might indicate the presence of a ring unresolved at the spatial resolution of the observations. Interestingly, this structure is slightly more prominent in the South-East side of the disk, as shown in Fig.~\ref{fig:rad_prof}, right, suggesting that it is azimuthally asymmetric. Other disks have shown asymmetric structure reminiscent of streamers or filaments trailing a major symmetric dust ring \citep{isella2018,perez2019}, but the spatial resolution of the observation of LkCa~15 impedes a similar characterization of the morphological feature. Both disks show a shallower tail in the outer regions, which has been already observed in LkCa~15 and other transition disks \citep[e.g.,][]{pinilla2018,jin2019}, in agreement with radial drift models predicting a smaller maximum grain sizes in the outer regions of the disk, leading to lower opacities at mm wavelength. In both cases, this extended emission shows a shoulder (meaning an intensity profile with negative second derivative), which may be tracing a shallow ringed structure at large radii. These outer shoulders are located at $\sim175\,$au and at $\sim63\,$au in LkCa~15 and J1610, respectively. 

\begin{figure*}[t]
\begin{center}
\includegraphics[width=0.48\textwidth]{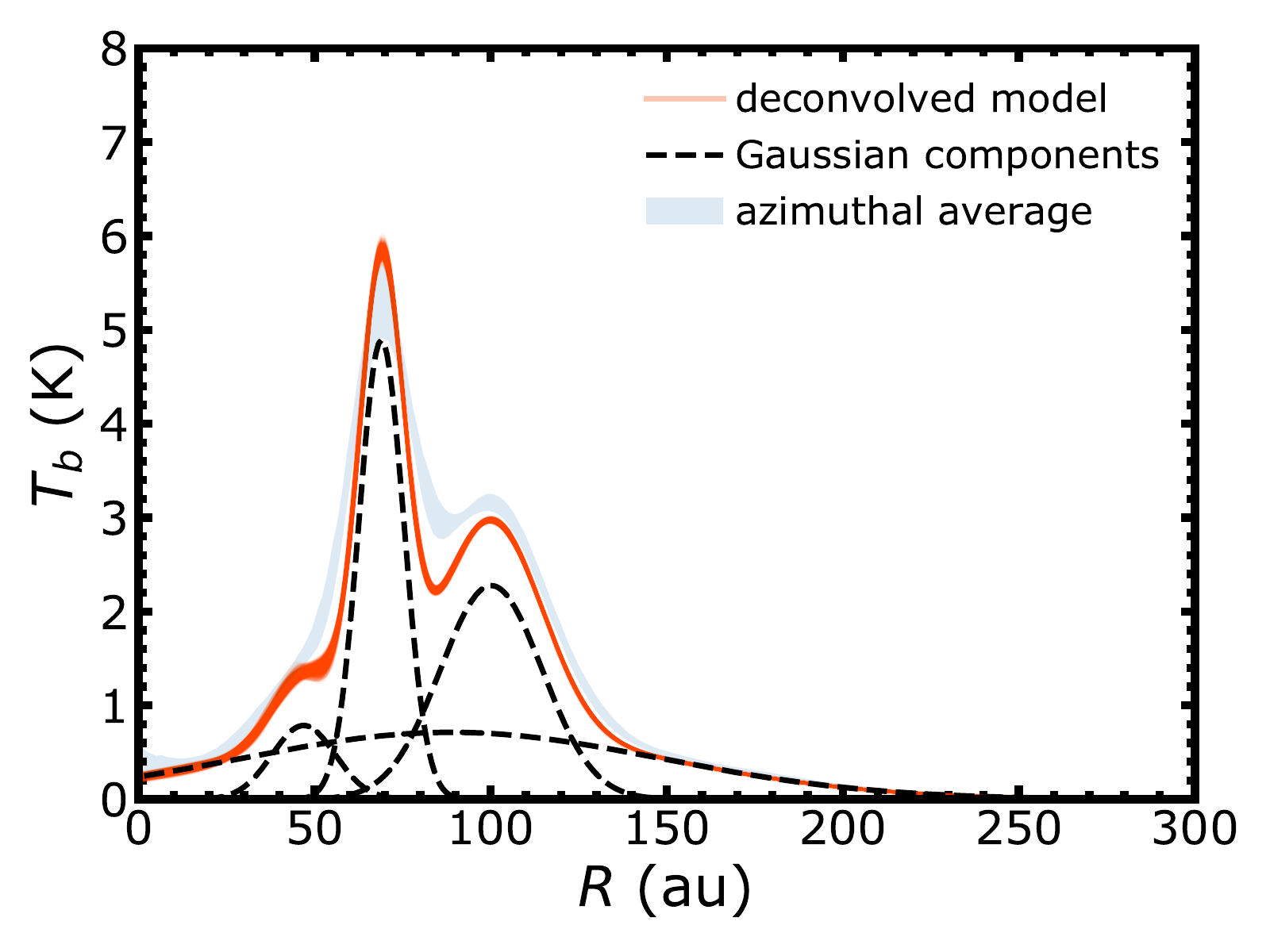}
\includegraphics[width=0.48\textwidth]{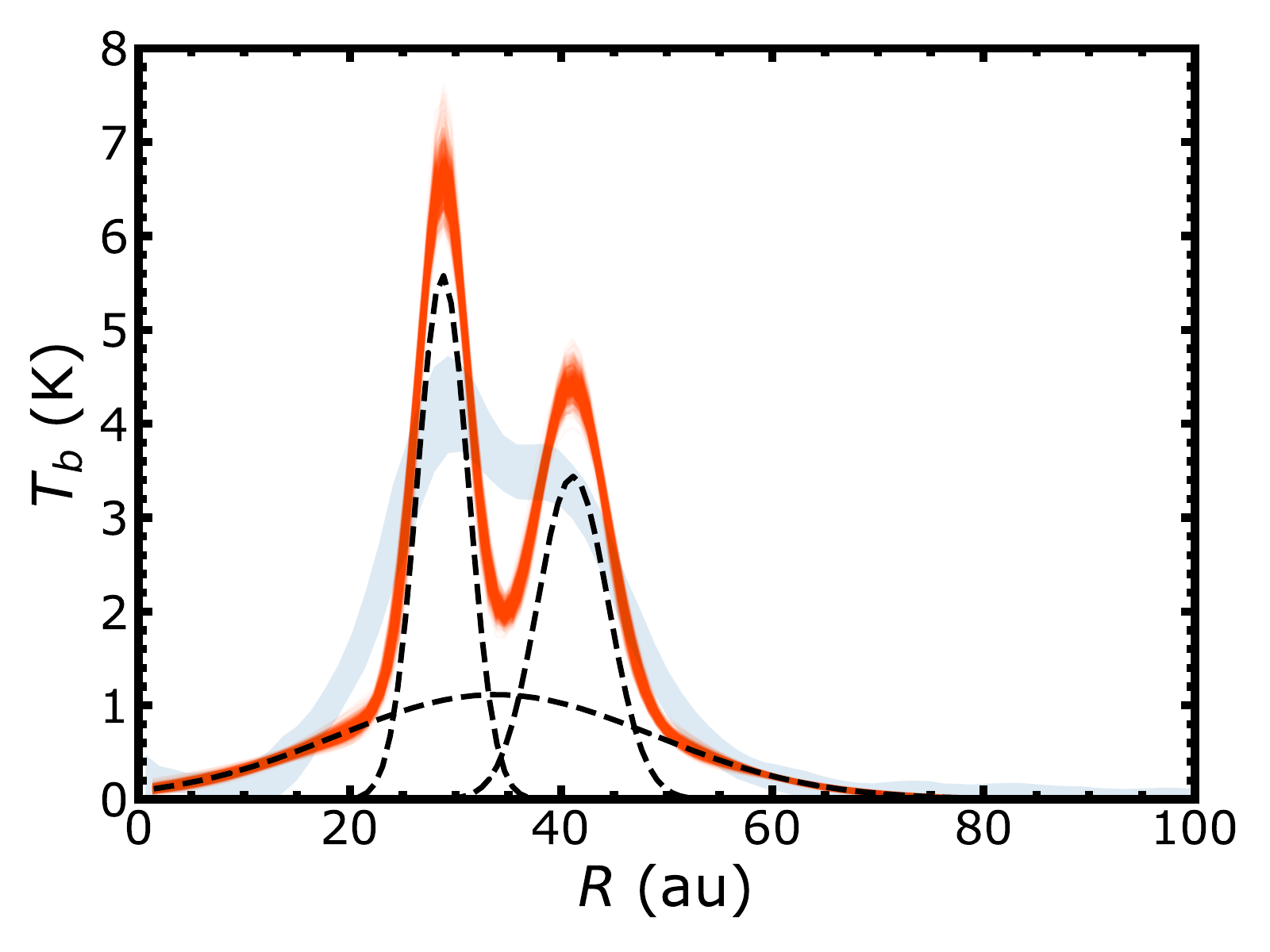}
\end{center}
\caption{Deconvolved brightness temperature radial profiles of LkCa~15 (left panel) and J1610 (right panel), calculated in the Rayleigh-Jeans approximation (see Sec.~\ref{sec:galario}). The red curves show 800 random realizations of the posterior distribution of the {\tt GALARIO} model, while the dashed black lines indicate the best fit of the Gaussian components used in the model. The underlying blue shaded region shows the deprojected and azimuthally averaged brightness temperature profile derived from Fig.~\ref{fig:maps} (see Sec.~\ref{sec:images}).}
\label{fig:rad_prof_galario}
\end{figure*}

\begin{table*}[t]
\centering
\caption{Median of the marginalized posteriors of the fitted parameters for LkCa~15, with associated statistical uncertainties from  the 16th and 84th percentiles of the marginalized distributions. The center of the emission is computed with respect to the phase center of the visibilities, i.e., RA (J2000) = 04:39:17.8059 and Dec (J2000) = +22:21:03.048.}
\begin{tabularx}{0.9\textwidth}{l c*{4}{Y}}
\toprule
Disk geometry &	&	&	&	&	\\
\midrule
$i$	&	$(^{\circ})$	& $50.16\pm0.03$	&	&	&	\\
PA	&	$(^{\circ})$	& $61.92\pm0.04$	&	&	&	\\
$\Delta$RA	&	(mas)	& $3.6\pm0.2$ 	&	&	&	\\
$\Delta$Dec	&	(mas)	& $-6.0\pm0.2$	&	&	&	\\
\midrule
Ring parameters	&	&	B47	&	B69	&	B100	&	Broad component \\
$I_i$	&	($\log\,$Jy/steradian)		& $9.078^{+0.018}_{-0.019}$	& $9.866^{+0.004}_{-0.005}$	& $9.535^{+0.003}_{-0.003}$ 	& $9.029^{+0.009}_{-0.010}$	\\
$R_i$	&	(au)		& $47.33^{+1.53}_{-1.20}$	& $68.99^{+0.17}_{-0.15}$	& $100.11^{+0.19}_{-0.19}$	& $88.92^{+1.53}_{-1.37}$	\\
$\sigma_i$	&	(au)		& $9.33^{+1.28}_{-1.11}$	& $6.32^{+0.16}_{-0.16}$	& $14.51^{+0.21}_{-0.21}$	& $59.87^{+0.54}_{-0.57}$	\\
\bottomrule
\end{tabularx}
\label{tab:LkCa15}
\end{table*}

\begin{table*}[t]
\centering
\caption{Same as Tab.\,\ref{tab:LkCa15} for J1610 and using RA (J2000) = 16:10:05.0032 and Dec (J2000) = -21:32:32.378.}
\begin{tabularx}{0.75\textwidth}{l c*{3}{Y}}
\toprule
Disk geometry &	&	&	&	\\
\midrule
$i$	&	$(^{\circ})$	& $37.80\pm0.20$	&	&	\\
PA	&	$(^{\circ})$	& $60.51\pm0.34$	&	&	\\
$\Delta$RA	&	(mas)	& $-0.9\pm0.4$ 	&	&	\\
$\Delta$Dec	&	(mas)	& $-1.8\pm0.3$	&	&	\\
\midrule
Ring parameters	&	&	B29	&	B41	&	Broad component \\
$I_i$	&	($\log\,$Jy/steradian)		& $9.925^{+0.019}_{-0.019}$	& $9.714^{+0.017}_{-0.019}$	&  $9.225^{+0.064}_{-0.067}$	\\
$R_i$	&	(au)		& $28.79^{+0.10}_{-0.10}$	& $41.11^{+0.13}_{-0.15}$	&  $33.81^{+0.50}_{-0.61}$	\\
$\sigma_i$	&	(au)		& $2.42^{+0.16}_{-0.16}$	& $3.36^{+0.24}_{-0.24}$	&  $15.10^{+1.04}_{-0.90}$	\\
\bottomrule
\end{tabularx}
\label{tab:J1610}
\end{table*}

\paragraph{Emission within the cavity.} The cavities of both disks present low levels of emission (see Figs.~\ref{fig:rad_prof} and \ref{fig:rad_prof_log}). In the following, we refer to an uncertainty driven by the nominal rms given in Sec.~\ref{sec:data}, and estimated in the outer regions of the maps. We note that it might underestimate the rms in the inner regions, for which deconvolution errors are significant due to the high dynamic range. J1610 does not show clear centrally-peaked emission, with an average surface brightness within the cavity of $\sim2\sigma$. LkCa~15 also shows a low surface brightness distributed within the whole cavity, but with clear inner disk unresolved emission located at the center, with a peak intensity of $73.7\pm6.9\,\mu$Jy\,beam$^{-1}$, reaching a S/N$\sim10.7\sigma$. The total flux within an area slightly larger than one beam (with same PA, and major and minor axes 1.2 larger than the ones of the beam) leads to a flux density of $58.9\pm7.2\,\mu$Jy. \citet{isella2014} also detected a point source at the center of LkCa~15 with VLA observations at 7\,mm with 70\,mas resolution, with a flux density of $16.6\pm3.6\,\mu$Jy. Taking into account a 10$\%$ uncertainty in the flux calibration of both the VLA and the ALMA observations presented in this paper, and assuming that the errors are Gaussian, we obtain a spectral index of $0.9\pm0.2$ between 1.35 and 7\,mm. This low value is incompatible with emission due to dust thermal radiation alone, and suggests that at 7\,mm there may be a contribution from either free-free or synchrotron emission by the ionized gas in the proximity of the central star \citep{panagia1975,reynolds1986}.

\begin{figure*}
\begin{center}
\begin{tabular}{c}
\includegraphics[width=0.9\textwidth]{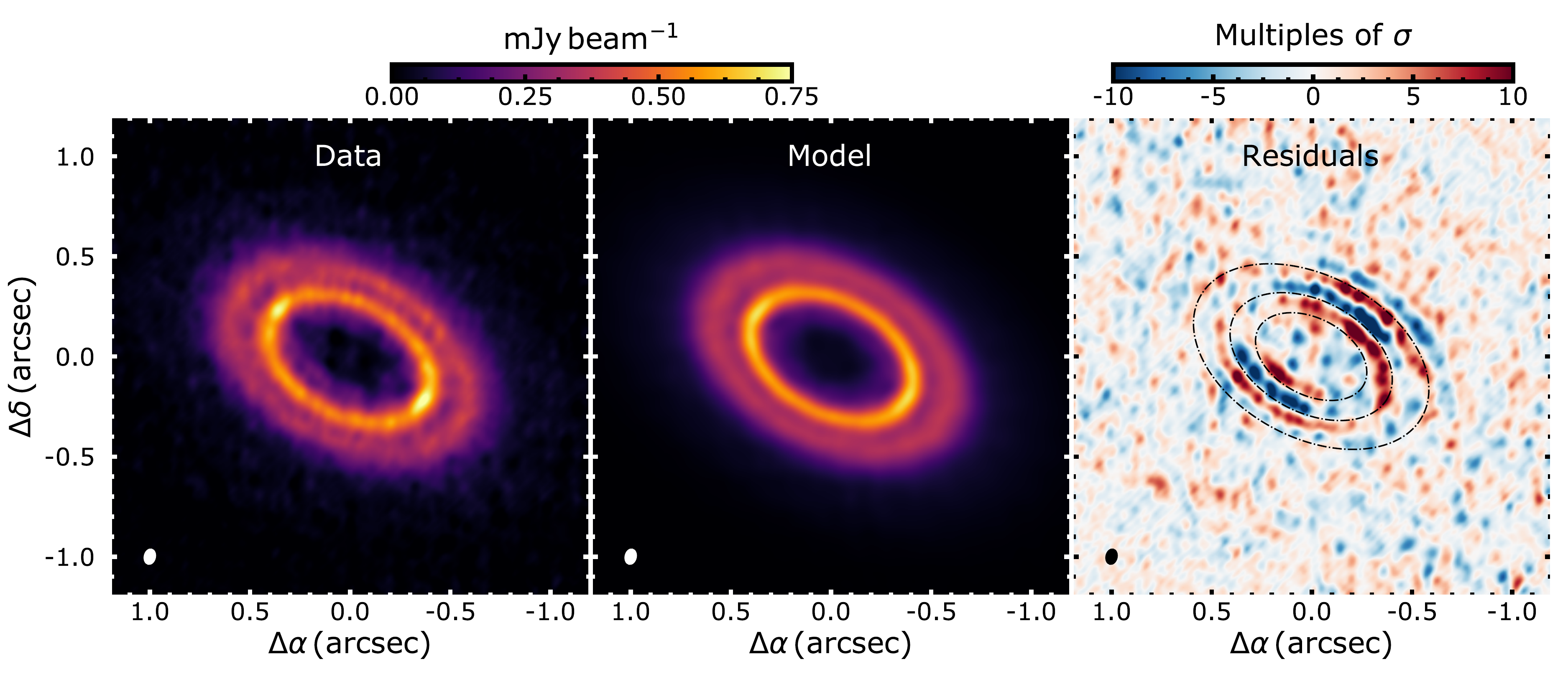}\\
\includegraphics[width=0.9\textwidth]{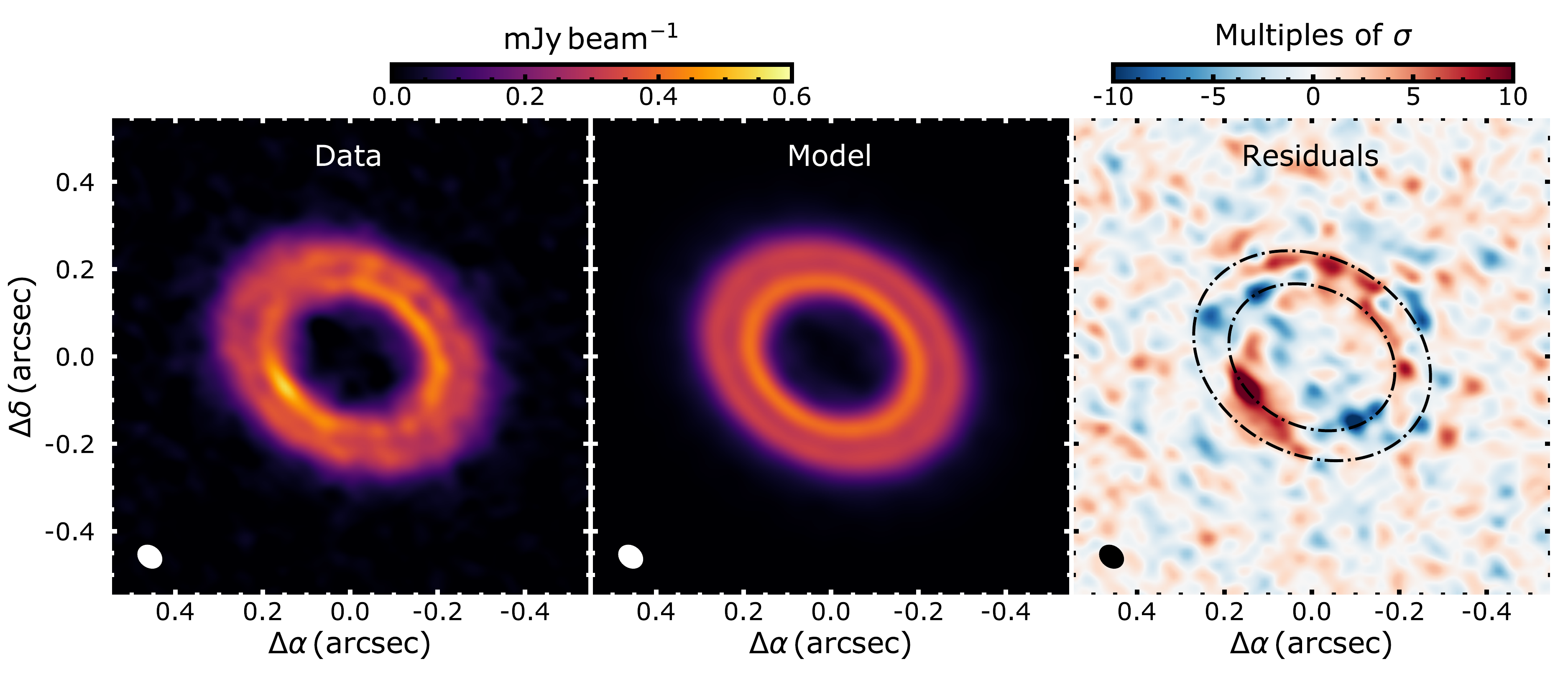}
\end{tabular}
\end{center}
\caption{From left to right: intensity map of the data, best fit model and residual visibilities for LkCa~15 (top) and J1610 (bottom). The intensity map of the best fit model model and residuals have been generated using the same uv-sampling as the original data. In the residual map, the black dashed-dotted lines show the location of the rings (see Tables~\ref{tab:LkCa15} and \ref{tab:J1610}).}
\label{fig:map_mod_res_lkca15}
\end{figure*}

The LkCa~15 image additionally presents a localized, unresolved, bright emission within the cavity, at a radius of $\sim0.22\arcsec$ ($\sim35\,$au) and along a position angle of $\sim236^\circ$ (see Fig.~\ref{fig:blob_lkca15}, left).  
The overbrightness is seen when imaging the disk with different weightings, in particular with robust parameter $<0.3$ that allows to separate it from the inner bright mm ring. To verify whether it is an imaging artefact, we measure the peak intensity at the location of the feature when imaging the disk with different weighting schemes, in particular with Briggs weighting and robust parameter ranging from 0.5 to -0.5. The results are shown in Fig.~\ref{fig:blob_lkca15}, right panel. The peak intensity varies with imaging parameters, ranging between $112\pm21$ to $211\pm14\,\mu$Jy\,beam$^{-1}$ (for robust parameter -0.5 and 0.3, leading to a nominal S/N of $\sim5.3$ and $\sim15\sigma$, respectively). This suggests that the feature is likely an imaging artefact, since the peak intensity of an unresolved point source should not vary with resolution element. However, considering the proximity of the over-intensity to the bright mm ring, we expect the peak intensity to decrease with increased angular resolution as they are better separated. We discuss this further in the uv-plane analysis in Sec.~\ref{sec:galario}. This discussion highlights the challenges of detecting faint point sources within cavities of transition disks, due to the high dynamic range and image reconstruction artefacts. Deeper observations with different array configurations and possibly different frequencies are needed to reach a final conclusion about the nature of these faint features, which can be interpreted as circumplanetary disks. Even more convincing cases, as the one of PDS~70c \citep{isella2019}, still need confirmation with new observations.

\subsection{Modelling of the continuum emission}
\label{sec:galario}
In order to better characterize the continuum emission described in Sec.~\ref{sec:images}, we model the intensity maps in the $(u,v)$-plane. We describe the continuum brightness distribution of LkCa~15 and J1610 by fitting the visibility points using an axisymmetric parametric model that consists in a set of radially Gaussian rings. The brightness profile of each model follows: 
\begin{equation}
I(R) = \displaystyle\sum_{i=1}^{N} I_i \; e^{-(R-R_i)^2/2\sigma_i^2}
\end{equation}
where $N$ is the number of rings considered, $I_i$ is the peak intensity at radius $R_i$,  $\sigma_i$ is the ring width. As each ring has three free parameters ($I_i$, $R_i$ and $\sigma_i$), the model has 3$\times N$ free parameters, plus four global disk parameters that are fitted for all rings together: the disk inclination ($i$), position angle (PA) and offset from the phase center ($\Delta$RA, $\Delta$Dec). Inspired by the image analysis (Sec.\,\ref{sec:images}), the model consists in a broad ring and 3 or 2 narrow rings for LkCa~15 and J1610, respectively, and is described with 16 and 13 free parameters, respectively. Flat priors are used for all parameters, with the constraint that one of the Gaussian rings has $\sigma_i>0.15\arcsec$, in order to reproduce the broad component at large radii. For each set of parameters, the Fourier transform of the model image is computed and sampled in the $(u,v)$ points of the dataset using {\tt GALARIO} \citep{galario}.  We sample the posterior distribution of our parameters with the \texttt{emcee} package \citep{emcee} with 160 walkers over 10000 steps after 2000 steps of burn-in. 

Tables \ref{tab:LkCa15} and \ref{tab:J1610} report the parameters of the maximum likelihood model for both targets, as well as statistical uncertainties for each parameter estimated from the 16$^{th}$ and 84$^{th}$ percentiles of the marginalized distributions. The error estimates have been computed after thinning the chains to avoid dependant samples in the posterior distributions. The brightness distribution of LkCa~15 is well represented by 3 narrow Gaussian rings centered at $\sim$47, $\sim$69 and $\sim$100\,au (with widths between $\sim$6 and $\sim$14\,au) and a broad component centered at $\sim$89\,au with a width of $\sim60$\,au. In J1610, a similar morphology leads to a good fit, with two narrow $\sim$2-3\,au-wide Gaussian rings located at $\sim$29 and $\sim$41\,au and a broad component with a width of $\sim15\,$au. The similar width of the two Gaussian rings in J1610 suggests that the estimated width of the rings is limited by the angular resolution of the observations (the FWHM of the beam along the minor axis is $\sim2\sigma_i$). The Gaussian curves composing the best-fit models are shown in dashed lines in Fig.~\ref{fig:rad_prof_galario}.

Figure\,\ref{fig:map_mod_res_lkca15} provides the images of the data, model and residuals for the best-fit model, synthesized with the same imaging parameters. The comparison between the observed visibilities and the model as a function of $(u,v)$ distance are shown in the Appendix (Fig.~\ref{fig:visibilities}). While the models reproduce well the overall morphology of the emission for both targets, residuals are found at high spatial frequencies. 
In LkCa~15, at the 10$\sigma$ level, residuals are found slightly inwards and outwards of the rings indicating a more complex morphology than prescribed with a Gaussian. The width of the rings in LkCa~15 is larger than the resolution element, indicating that the rings are partly resolved and deviations from a Gaussian profile can be detected with these observations. Interestingly, both negative and positive residuals are mostly seen along the minor axis, which indicates that the intensity prescription is a good representation of the data where the linear resolution is maximum (along the major axis). The residuals along the minor axis can be explained by projection and radiative transfer effects with a finite ring vertical thickness \citep[see also discussion in][]{huang2020}. The NW over-brightness in both B69 and B100 is apparent in the residuals, as is the azimuthal asymmetry in the inner shoulder (B47). This confirms the analysis in the image plane, with the shoulder being more prominent in the SE side of the disk (Fig.~\ref{fig:rad_prof}, right). In J1610, the largest residual is azimuthally broad and located in the SE. Interestingly, the positive and negative residuals are mostly located along almost perpendicular directions (minor and major axis, respectively) which could be related to shadowing by the inner disk, as seen in DoAr~44 \citep{casassus2018}. In both disks, the outer shoulder at $\sim175\,$au and $\sim63\,$au is not seen in the residuals (nor in their azimuthal average), with the extended Gaussian component reproducing the intensity profile within $3\sigma$.  Deeper observations might be needed to test whether they can be attributed to shallow rings at large distance from the star.

Within the central cavities, no residuals are found above the $4\sigma$ level. The models shown in Fig.~\ref{fig:rad_prof_galario} both present low surface brightness emission within the cavities. Such prescription works well in both cases, even though the S/N within the cavities is low enough in both disks that other functional forms could have been equally representative of the intensity distribution. In LkCa~15, the point like feature discussed in Sec.~\ref{sec:images} is not recovered in the residuals above $4\sigma$, suggesting again that this is due to an imaging artefact driven by the high dynamic range of the observations.

\section{Comparison with hydrodynamical simulations}
\label{sec:hydro}

In this section, we aim to study the conditions for the formation of multiple rings in transition disks that host an inner dust-depleted cavity. We assume that the rings are caused by the interaction with a planet located at the pressure maximum at the edge of the central cavity. We use the better resolved LkCa~15 as a test case. We stress that our goal is not to reproduce a perfect match with the observations, since this would not imply that the solution is unique and would require fine-tuning of the hydrodynamical simulations while lacking possibly relevant physical mechanisms.

\subsection{Set-up}
\label{sec:set-up}
We adopt a parameterized initial gas surface density profile based on the analysis of low-resolution $^{12}$CO ($J$=3-2) observations of LkCa~15 by \citet{jin2019}, which well reproduces the inner cavity imaged at lower angular resolution. The profile we use follows
\begin{equation}
\label{eqn:sigma}
\Sigma_{\rm g, init}(R)   =  \Sigma_c \left(\frac{R}{R_c}\right)^{-\eta} \arctan\left[ \left( \frac{R}{R_c} \right)^{\psi} \right],
\end{equation}
where $R_c = 65$~au, $\eta=4$, and $\psi = 10$. We choose $\Sigma_c = 41~{\rm g~cm}^{-2}$ so that the total gas disk mass inside of 600~au is $0.1~M_\odot$. We point out that our choice of $R_c=65$~au differs from the best-fit value of \citet{jin2019}, $R_c=45$~au. We need to adopt $R_c=65$~au because grains are aerodynamically dragged towards the pressure peak near $R=R_c$ and form a ring there; with $R_c=45$~au, hydrodynamical simulations show that a ring forms at a too-small radius compared with the continuum observation. 

Using the surface density profile described as in Eq.\,(\ref{eqn:sigma}), we create the initial disk temperature by running iterative Monte Carlo radiative transfer (MCRT) calculations using RADMC-3D \citep{radmc3d}, following the method described in Appendix A of \citet{bae2019}. To briefly summarize, this iteration finds self-consistent three-dimensional density and stellar irradiation-dominated temperature structures that  satisfy the hydrostatic equilibrium. We assume $10^{-3}~M_\odot$ of total dust mass distributed among grains with sizes between $0.01~\mu$m and 1~cm, adopting a power-law size distribution with power-law exponent of $-3.5$. We further assume that small dust grains having sizes between 0.01 and 1~$\mu$m are perfectly coupled with disk gas and determine the disk temperature profile. This small grain component has total $9\times10^{-6}~M_\odot$ with the assumed power-law size distribution. For the small grain composition, we assume that they are compact monomers consist of $60~\%$ silicate and $40~\%$ amorphous carbon, having an internal density of $2.7~{\rm g~cm}^{-3}$. We adopt optical constants of silicate and amorphous carbon from \citet{draine84} and \citet{li97}, respectively. We assume a stellar radius of $1.6~R_\odot$ and an effective temperature of 4500~K \citep{donati2019} for the MCRT iterations. Each iteration runs with $10^9$ photon packages.

The resulting density-weighted, vertically integrated temperature from the MCRT iterations is presented in the Appendix in Fig. \ref{fig:disk_temperature}. Note that the disk temperature drops rapidly around the location of the density peak at 65~au as stellar photons are scattered/absorbed at the wall outside of the inner cavity. Adopting a stellar mass of $1.25~M_\odot$ \citep{donati2019} and a mean molecular weight of 2.4, the disk aspect ratio $H/R$ between 30 and 300 au can be well described by the following functional form: 
\begin{equation}
\label{eq:temperature}
{H \over R} = 0.082 \left( {R \over R_c} \right)^{0.375} + 0.022\left[ 1 + \tanh \left(- {R-50~{\rm au} \over 15~{\rm au}} \right) \right].
\end{equation}

Using the surface density and temperature profiles described in Eq. \ref{eqn:sigma} and \ref{eq:temperature}, we obtain initial azimuthal velocity that satisfies hydrostatic equilibrium. The initial radial velocity is set to zero.

\subsection{Hydrodynamical simulations}

We carry out two-dimensional hydrodynamical simulations using the Dusty FARGO-ADSG code \citep{baruteau2019}. This is an extended version of the publicly available FARGO-ADSG \citep{masset00,baruteau08a,baruteau08b}, with Lagrangian test particles implemented \citep{baruteau16}. The simulation domain extends from 20 to 300~au in the radial direction and covers the entire $2\pi$ in azimuth. We adopt 672 logarithmically-spaced grid cells in the radial direction and 1556 uniformly-spaced grid cells in the azimuthal direction. With this choice, one scale height at the location of the planet is resolved with 22 grid cells in both radial and azimuthal directions. A wave-damping zone \citep{devalborro06} is implemented at both inner and outer radial boundaries. The disk viscosity $\nu$ is characterized by a uniform viscosity parameter $\alpha=10^{-4}$, where $\nu = \alpha c_{\rm s} H$, with $c_{\rm s}$ being the local sound speed.

We insert a 60 Earth-mass planet at 70~au, at the radial location of the middle continuum ring in the LkCa~15 disk, assuming a fixed circular orbit. This planetary mass has been chosen after running a coarse grid of different planet masses: 30, 60, and 90~$M_\oplus$. In short, with the viscosity considered in the model,  a $30~M_\oplus$ planet does not create any rings and gaps, whereas a $90~M_\oplus$ planet creates a qualitatively similar morphology to a $60~M_\oplus$ planet in terms of the number of rings and gaps, while the contrast between rings and gaps is much larger. A more general discussion about the model degeneracy can be found in Sec.~\ref{sec:discussion}. We note that we are agnostic about the origin of the inner cavity and we do not introduce additional planets in the inner cavity, since this would introduce a high number of additional free parameters.

\begin{figure*}
    \centering
    \includegraphics[width=\textwidth]{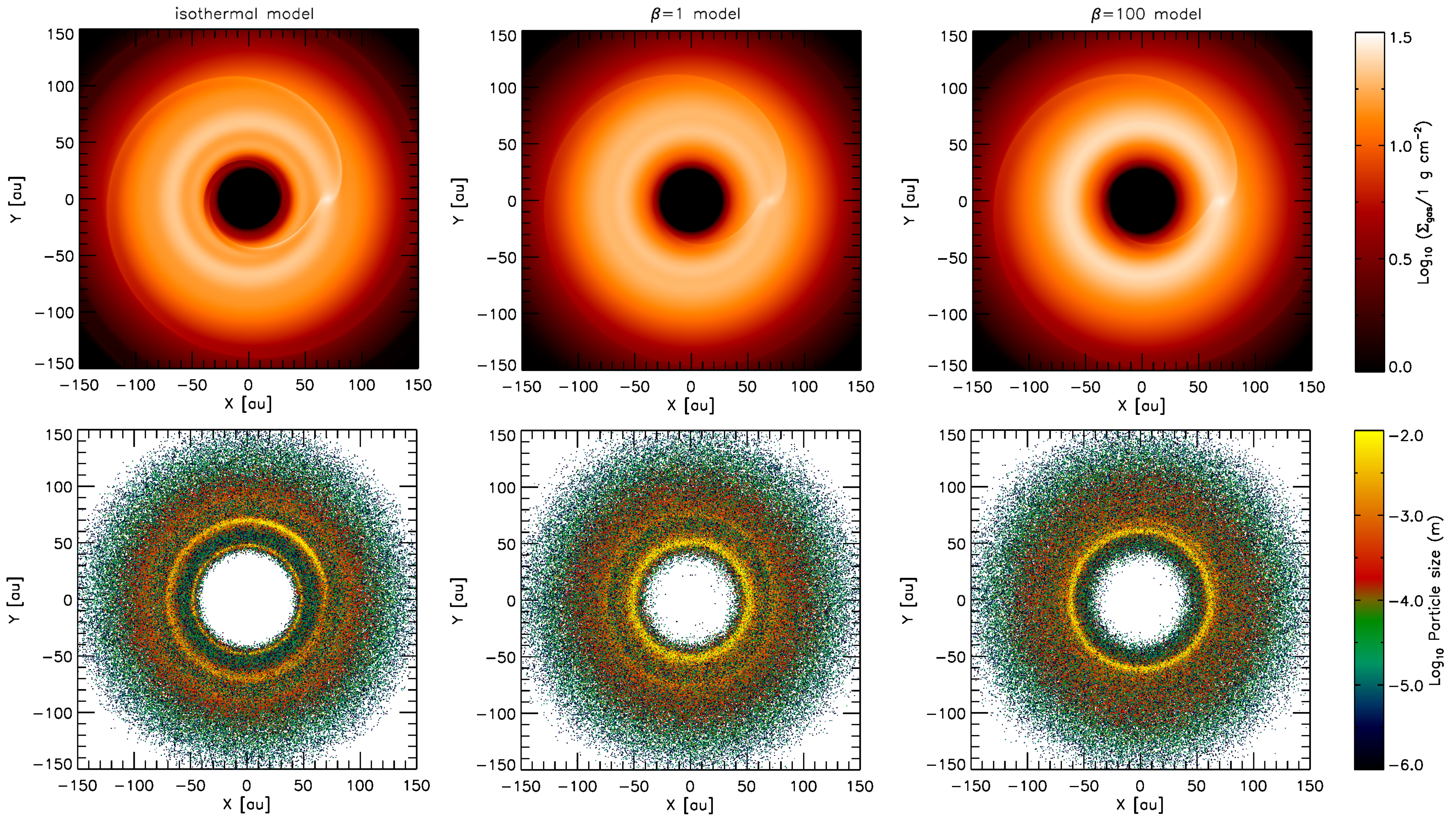}
    \caption{Two-dimensional distributions of (upper panels) gas and (lower panels) particles. From left to right: (left) isothermal model, (middle) $\beta=1$ model, and (right) $\beta=100$ model. The planet is located at (X, Y) = (70~au, 0~au). The disk and planet orbit clockwise about the central star.}
    \label{fig:densxy}
\end{figure*}

\begin{figure*}
    \centering
    \includegraphics[width=\textwidth]{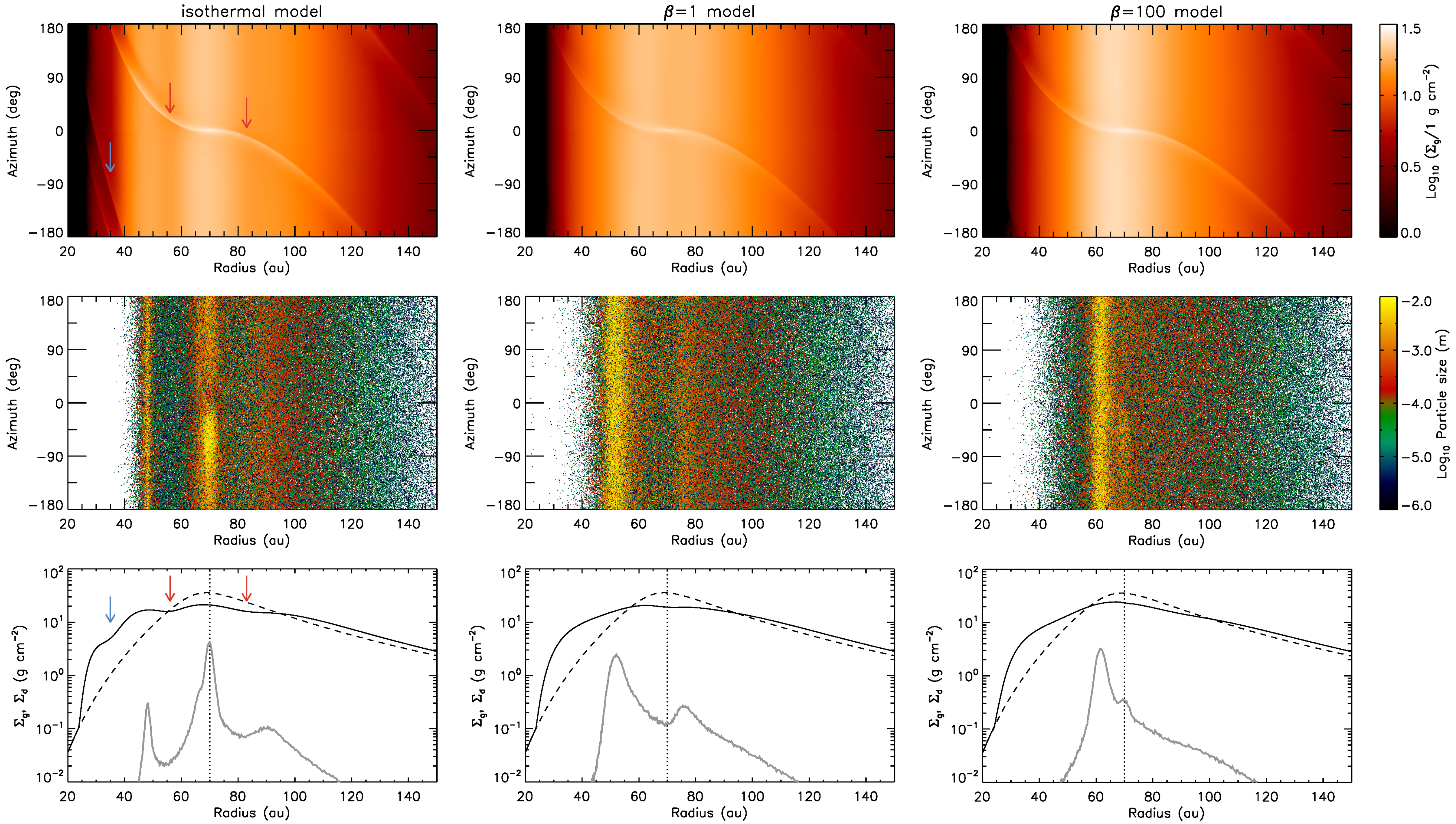}
    \caption{Two-dimensional distributions of (upper panels) gas and (middle panels) particles in a radius - azimuthal angle plane. The lower panels show azimuthally-averaged radial profiles of gas (solid black lines) and dust (solid gray lines) surface density. The black dashed curves present the initial gas surface density profile, while the black dotted lines present the radial location of the planet. In the isothermal model, the red and blue arrows indicate the location of the shocks due to the  primary (red) and secondary (blue) spiral arms. From left to right: (left) isothermal model, (middle) $\beta=1$ model, and (right) $\beta=100$ model. The planet is located at $(R, \phi)$ = (70~au, $0^\circ$).  The disk rotation is upward in the upper and middle panels.}
    \label{fig:densrphi}
\end{figure*}

\begin{figure*}
    \centering
    \includegraphics[width=\textwidth]{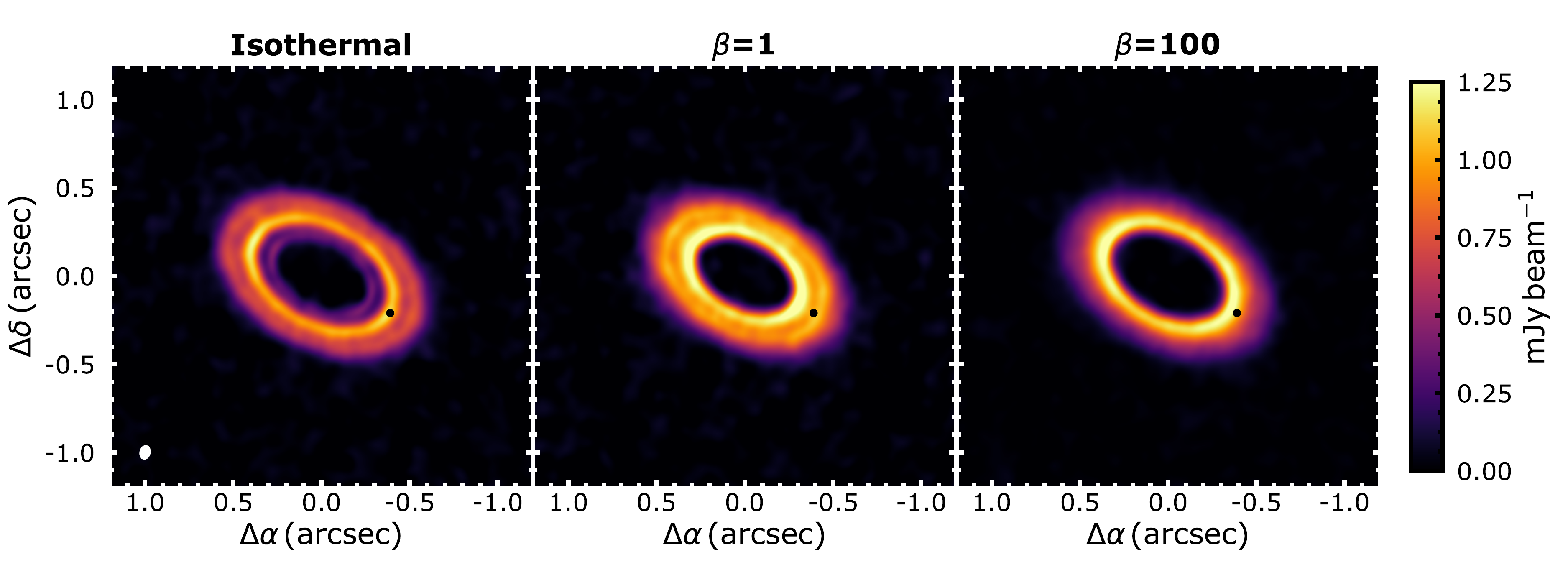}
    \caption{Simulated continuum images based on (left) isothermal, (middle) $\beta=1$, and (right) $\beta=100$ models. The location of the planet $(-0.39'', -0.21'')$ is shown with a black filled circle in each panel. The synthesized beam is shown with a white ellipse in the lower-left corner of the left panel. Note the diverse continuum morphology that various treatments of the disk thermodynamics can produce.}
    \label{fig:simulated_obs}
\end{figure*}

Simulations are run for 1000 planetary orbits ($\simeq$0.52~Myr), by which time the gas and particle distribution reached a quasi-steady state. After this period, we insert $3\times10^5$ Lagrangian test particles between 50 and 150~au, with a uniform dust-to-gas mass ratio of $1~\%$ across this radial region, and we run simulations for additional 300 planetary orbits. The size of the test particles is uniformly distributed between 1~$\mu$m and 1~cm in logarithmic space. The mass of each test particle is determined such that, at each radius in the disk, the dust mass is distributed over a range of dust sizes, from 1~$\mu$m and 1~cm to have the mass per interval in $\log(s)$ be proportional to $s ^{0.5}$, consistent with a dust size distribution of $n(s) \propto s^{-3.5}$.
We assume a dust bulk density of $\rho_s = 1.26~{\rm g~cm}^{-3}$, which corresponds to the density of aggregates with $30~\%$ silicate matrix and $70~\%$ water ice. The bulk density is lower than that of sub-$\mu$m grains responsible for the thermalization of the stellar radiation (see Sec. \ref{sec:set-up}), since for the small dust grains in the upper layers of the disk we do not expect ice-coating.

Recent numerical simulations of planet-disk interaction have shown that the propagation and dissipation of planet-driven spiral waves and the number and depth of associated gaps strongly depend on the thermal properties of the disk gas, in particular the cooling timescale of the disk gas  \citep{miranda2019, zhang2019,weber2019,ziampras2020}. Inspired by these studies, we run three hydrodynamical simulations: (1) a locally isothermal simulation, adopting an isothermal equation of state (hereafter isothermal model); (2) an adiabatic simulation, adopting an adiabatic equation of state with an adiabatic index $\gamma=1.4$ and $\beta=1$ (hereafter $\beta=1$ model), where $\beta$ is defined as the multiplication of the cooling time $t_{\rm cool}$ and local orbital frequency $\Omega$, $\beta \equiv t_{\rm cool} \Omega$; and (3) an adiabatic simulation with $\gamma=1.4$ and $\beta=100$ (hereafter $\beta=100$ model). In $\beta=1$ and $\beta=100$ models, the disk temperature is relaxed toward the initial temperature over the cooling timescale $t_{\rm cool}$.

\subsection{Model results}

In Fig.~\ref{fig:densxy}, we present two-dimensional disk surface density and particle distributions. Gas and particle distributions in a radius - azimuthal angle plane, along with azimuthally-averaged radial profiles of gas and dust surface density, are shown in Fig.~\ref{fig:densrphi}.

In the isothermal model, the inner and outer primary spiral arms (indicated with red arrows in Fig.~\ref{fig:densrphi}) create shocks at $\pm13$~au from the planet, which corresponds to about $\pm2$ scale heights at the radial location of the planet. This is in a good agreement with the shock distance predicted by linear theory \citep{goodman2001}. A secondary spiral arm forms in the inner disk (indicated with blue arrows in Fig.~\ref{fig:densrphi}) and creates shocks around 35~au. As these spiral arms form they generate shocks, opening gaps in the disk \citep{bae2017}. The gas pressure has local maxima between the gaps and beyond the outer gap at 48, 70, and 90 au, around which radii particles are aerodynamically dragged and trapped. Therefore, the planet generates three rings and two gaps in the dust disk, and the planet is embedded in the middle ring. In addition, within the co-rotation region, large particles ($\sim 1$~cm) are preferentially collected around one of the Lagrangian points, L5, which locates $60^\circ$ behind the planet, leading to azimuthal asymmetries in the density distribution of large grains.

In $\beta=1$ model, the planet opens a single, broad gap around its orbit. This distinct feature of single gap opening around a sub-thermal-mass planet is known to happen when $\beta \simeq 0.1 - 1$ as linear damping plays a more important role in angular momentum transfer than non-linear shock dissipation \citep{miranda2019,zhang2019}. As a result, two gas pressure peaks form in this model, one at the outer edge of the inner cavity at 52~au and the other beyond the gap at 75~au. The planet is embedded within the gap, sandwiched by two dust rings.

In $\beta=100$ model, the disk behaves nearly adiabatically and the perturbation driven by planet-driven spiral arms is much weaker compared with the isothermal model \citep{miranda2019,zhang2019}. A secondary spiral arm forms in the inner disk around $30$~au, but the density perturbation driven by the spiral arm is much weaker compared with that seen in the isothermal model. With a finite viscosity implemented in the simulation ($\alpha=10^{-4}$), neither primary nor secondary spiral arms open a gap. The gas pressure peaks slightly inward of the planet's orbit, at 62~au. Large particles ($\sim 1$~cm) are collected around Lagrangian point L5, similar to the isothermal model.

Based on the three models, we generate simulated continuum images. We first create 50 logarithmically-spaced grain size bins between $1~\mu$m and 1~cm and compute the surface density of each grain component at each simulation grid cell. We then expand the dust disk vertically, assuming that grains are in a hydrostatic equilibrium within isotropic turbulence characterized by $\alpha=10^{-4}$ and have a Gaussian distribution. With this assumption, the scale height of dust particle having size $s$, $H_d(s)$, can be written as
\begin{equation}
    H_d(s) = H \times {\rm min} \left( 1, \sqrt{{\alpha} \over {{\rm min} ({\rm St}, 1/2)(1+{\rm St}^2)}} \right),
\end{equation}
where St$\equiv \pi \rho_s s /2 \Sigma_{\rm g}$ is the Stokes number of the particle \citep{birnstiel2010}. The raw continuum emission is obtained with RADMC-3D \citep{radmc3d}, considering both absorption and anisotropic scattering using the Henyey–Greenstein approximation. We note that scattering dominates the continuum emission in the simulated observations, supporting recent explanations to low optical depths estimated in spatially resolved continuum observations \citep{zhu2019}.  We assume a distance of 158.9~pc and disk inclination and P.A. of $50^\circ$ and $62^\circ$, respectively. We then convolve the raw images with the same synthesized beam used for the LkCa~15 observation (i.e., 68$\times$47 mas with a P.A. of $-12.6^\circ$) and add random noise at the same level to the observation (i.e., $6.9~\mu$Jy~beam$^{-1}$).

The resulting simulated continuum images are presented in Fig.~\ref{fig:simulated_obs}. As shown, a $60~M_\oplus$ planet can produce diverse continuum morphology depending on the disk's cooling timescale. In the isothermal model, the planet generates three continuum rings separated by two gaps. The planet is embedded in the middle ring. In $\beta=1$ model, the planet opens a gap and two rings. The planet is embedded in the gap. In $\beta=100$ model, the disk appears as if it does not have any planets beyond the cavity; the disk is well characterized by the inner cavity and a single ring without footprints from the $60~M_\oplus$ planet.

\section{Discussion}
\label{sec:discussion}
\paragraph{Rings in transition disks.} About 40 disks that have been observed at moderate angular resolution ($\sim$0.2-0.4\arcsec{}) show evidence for dust depleted inner cavities with radius of a few tens of au \citep[e.g.,][]{pinilla2020}. A few of these disks have been observed at about ten times better resolution and present small scale substructure beyond their cavity \citep[e.g.,][]{kudo2018, perez2019,huang2020}. All of them host multiple rings, up to four \citep{perez2019}, except HD\,100453 that is a known binary system. In addition to rings, some targets show localized azimuthal asymmetries \citep{perezlm2018, dong2018}. Although the number of transition disks observed at high resolution is significantly lower than those of classical disks with no cavity \citep[e.g.,][]{andrews2018}, it seems that rings are the most common type of substructure also in transition disks as in classical disks \citep{huang2018b}. These rings are likely tracing  the trapping of dust grains \citep{dullemond2018}, possibly as a consequence from the presence of companions located inside the cavity. This enables to maintain a significant dust mass in the outer disk over few million years, as mm dust particles would otherwise rapidly drift towards the star \citep[e.g.,][]{brauer2008}. While most of the rings in transition disks can be well modelled with a radial Gaussian profile, some objects clearly show a strong asymmetry in their innermost ring, with an additional shoulder located inwards \citep[in LkCa~15 and HD\,169142; this paper,][]{perez2019} or outwards of the ring \citep[as in DM\,Tau and GM\,Aur;][]{kudo2018,huang2020}. Higher angular resolution (and possibly longer wavelengths to avoid optical thickness) will help determining the nature of these peculiar morphologies, which may be connected to the physical mechanism carving the prominent cavities in these disks. At the same time, outer rings show shallower profiles in the outward direction. This is the case for B100 and B41 in LkCa~15 and J1610, respectively. These radial asymmetries in the intensity profile of dust rings are of particular interest to discern what is causing them. Indeed gaps generated by planets are expected to create an asymmetric pressure profile, which would lead to skewed rings in the dust \citep[e.g.,][]{dullemond2018}. This is even more pronounced when different grain sizes of dust are considered in the modelling, with large grains well confined at the pressure maximum but smaller grains tracing the underlying gas surface density \citep[e.g.,][]{pinilla2018}. \citet{pinilla2018} also showed how this effect can depend on the stellar mass. A larger sample of transition disks spanning different stellar and disk properties imaged at high angular resolution would help understand the physical origin of this common feature in the intensity profiles.

\begin{figure}
    \centering
    \includegraphics[width=\columnwidth]{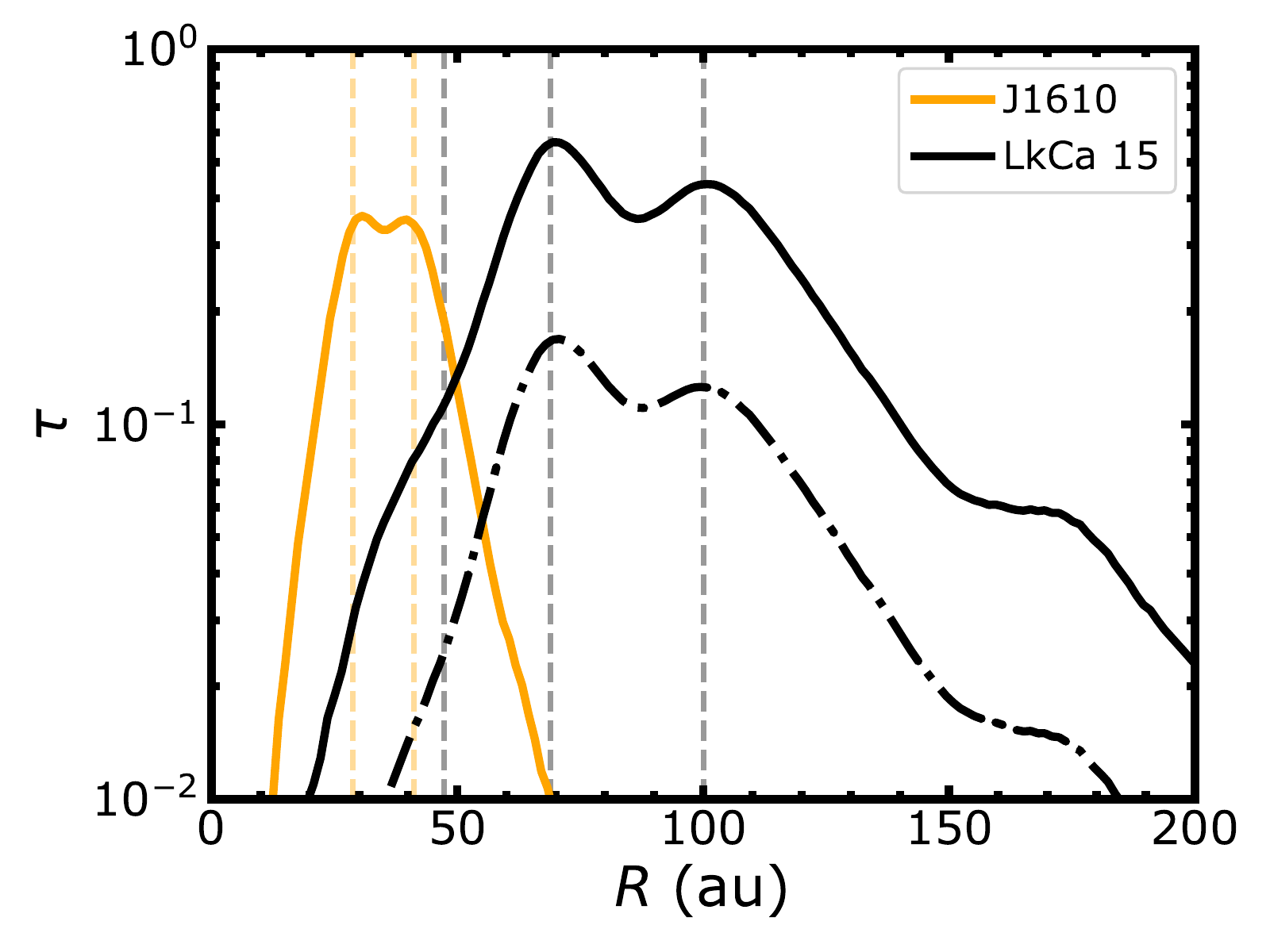}
    \caption{Optical depth profiles of both disks derived from Eq.~\ref{eq:tau}. The solid curves show the optical depth profiles assuming the temperature profile from Eq.~\ref{eq:temp}. The dashed-dotted line portrays the optical depth profile of LkCa~15 obtained with the temperature profile calculated with the radiative transfer model (see Sec.~\ref{sec:hydro} and Eq.~\ref{eq:temperature}). The optical depth profiles have been computed assuming absorption opacities only. The vertical dashed lines indicate the rings locations as derived from the $(u,v)$-plane modelling.}
    \label{fig:optical_depth}
\end{figure}

\paragraph{Optical depth.} From the intensity profiles, we compute optical depth profiles under the assumptions of a uniform opacity across the disks, and that the albedo of the dust grains is negligible. The optical depths have been computed from:
\begin{equation}
\label{eq:tau}
    I_\nu(R) = B_\nu(R) (1-e^{-\tau_\nu(R)}),
\end{equation}
with $B_\nu$ being the Planck function at the frequency of the observations, and $I_\nu$ being the azimuthally averaged intensity profile derived in Sec.~\ref{sec:results}. For LkCa~15 two midplane temperatures are used in the calculation, the one computed with the radiative transfer model discussed in Sec.~\ref{sec:hydro} (see Eq.~\ref{eq:temperature}), and one expression used to derive optical depth profiles in the DSHARP large program \citep{andrews2018,huang2018b}:

\begin{equation}
T(R)= \left( \frac{\varphi L_{*}}{8\pi R^2\sigma_{\rm SB}} \right)^{0.25},
\label{eq:temp}
\end{equation}
where $\sigma_{\rm SB}$ in the Stefan-Boltzmann constant, and $\varphi$ is the flaring angle, which we set equal to $0.02$ as in \citet{huang2018b}. Only this temperature profile has been used for J1610. The rings appear as marginally optically thick, with peak values ranging between 0.18 and 0.6 (depending on the exact temperature structure) for LkCa~15, and being $\sim0.3$ for J1610. The optical depth profiles are shown in Fig.~\ref{fig:optical_depth}, where LkCa~15 is presented with two curves based on different temperature profiles.  A maximum optical depth of $0.3-0.6$ is remarkably similar to the ones obtained in other studies \citep{huang2018b,facchini2019}. This may be caused by the presence of very large grains with significantly high albedo, with the optical depth peak derived from Eq.~\ref{eq:tau} being a lower limit in the absence of scattering \citep[e.g.,][]{liu2019,zhu2019}. Indeed, it is scattering what dominates the simulated continuum images presented in Figure \ref{fig:simulated_obs}. Observations at multiple wavelengths (more than 2) are needed to estimate how important scattering can be at mm wavelengths \citep[e.g.,][]{carrasco2019,huang2020} and whether it can explain the observational result of roughly uniform maximum optical depth in rings and inner regions across different disks. \citet{stammler2019} instead proposed that a maximum optical depth of $\sim0.5$ can be driven by streaming instability self-regulating the dust-to-gas ratio within pressure maxima \citep[see also][]{dullemond2018}, which would be indirect evidence of planetesimal formation with the dust rings observed with ALMA. To disentangle between these different possibilities, multi-wavelengths observations at high spatial resolution are needed. 

\paragraph{Dust masses and planetesimal formation.} From the optical depth profiles, it is possible to compute the amount of dust present in the rings by knowing that $\Sigma_{\rm d}(R)=\tau_\nu/\kappa$. For a given temperature profile and a given opacity, these values represent a lower limit, since we assume negligible albedo of the dust grains. The largest uncertainties lay in the dust opacity estimates, with values being unknown within 1 dex from the value we assume. From the opacity laws used in Sec.~\ref{sec:hydro}, we have an absorption opacity of 2.6 cm$^{2}$\,g$^{-1}$ at 1.3\,mm. For LkCa~15 we compute the mass of B47 and B69 together, integrating the optical depth profile between 19 and 88\,au. As for B100, we compute the dust mass between 88 and 150\,au. By using the temperature profile of Eq.~\ref{eq:temperature}, we obtain dust masses of 51 and $182\,M_\oplus$ for B47+B69 and B100, respectively, whereas using the temperature profile of Eq.~\ref{eq:temp} leads to dust masses of 32 and $113\,M_\oplus$ for the same regions. We integrate the optical depth profile of J1610 between 9.7 and and 35.8\,au for B29 and between 35.8 and 72.2\,au for B41, obtaining 10 and $19\,M_\oplus$ respectively.

These disks are still retaining a large amount of dust mass in this phase of their evolution, and planetesimal formation within these radially confined regions may still be operational via streaming instability \citep[e.g.,][]{youdin2005,johansen2007}. In order to assess whether planetesimal formation via streaming instability may be occurring within the dust rings, we need to compare the gas with the dust surface density within the dust rings. We use the dust surface density computed from the optical depth profile, and we compare it with the upper limit of the gas surface density at the location of the dust rings. In order to have a physically motivated upper limit, we require that the disk is not locally gravitationally unstable, since this would develop spiral structures that we do not see in the ALMA images \citep[e.g.,][]{GI_review}:

\begin{equation}
    Q_{\rm Toomre} \equiv \frac{c_{\rm s}\Omega_{\rm K}}{\pi G \Sigma_{\rm g}}>2,
\end{equation}
where $c_{\rm s}$ is the local sound-speed, $\Omega_{\rm K}$ is the Keplerian angular velocity and $\Sigma_{\rm g}$ is the gas surface density. The sound speed is estimated from Eq.~\ref{eq:temp}. With these calculations, we find that for LkCa~15, the dust rings B69 and B100 show that $\Sigma_{\rm g}/\Sigma_{\rm d}<[68,62]$ at their intensity peaks, respectively. As for J1610, B29 and B41 present $\Sigma_{\rm g}/\Sigma_{\rm d}<[190,138]$. If local turbulence is driven by the streaming instability itself, planetesimals are expected to form whenever $\Sigma_{\rm g}/\Sigma_{\rm dust}\lesssim50$ \citep[e.g.,][]{bai2010}, in which case the upper limits we obtained for LkCa~15 are intriguingly close to this value. If instead an additional turbulence component is present, the vertical settling of dust grains is reduced and the nominal condition for planetesimal formation to occur is less stringent \citep[e.g.,][]{gole2020}. Note, however, that the co-existence of vertical turbulent motions as driven by other instabilities as the vertical shear instability or magneto-rotational instability may promote rather than suppress planetesimal formation \citep[][]{johansen2007,schafer2020}.

A possible way to derive information about the gas turbulence in disks is to determine how radially (or vertically) confined dust grains are in disks. In the following, we look at the radial concentration of dust particles at the rings locations in the assumption that these are dust traps. We follow the arguments of \citet{dullemond2018} to characterize radially resolved rings in the DSHARP sample. First of all, we compute how radially confined the dust rings are, in order to have a lower limit on the combination of the turbulence viscosity $\alpha$ and the characteristic grain size within the rings. \citet{dullemond2018} showed that:

\begin{equation}
    \frac{\alpha}{{\rm St}} = \left[ \left( \frac{\sigma}{\sigma_{\rm d}} \right)^2-1 \right]^{-1},
\end{equation}
where ${\rm St}$ is the Stokes number of the characteristic dust grain size entrained in the ring (we can think at the characteristic size as the one dominating the emission at the observed wavelength), and $\sigma$ and $\sigma_{\rm d}$ are the widths of the gas and dust components of the rings, approximated as radial Gaussian profiles in the proximity of the rings. While \citet{dullemond2018} computed the dust Gaussian widths by fitting the rings from the imaged radial intensity profiles, we use the ones obtained from the $(u,v)$-plane fitting (values are reported in Tables~\ref{tab:LkCa15}). As for the gas Gaussian widths, we determine lower and upper limits using the same arguments as in \citet{dullemond2018}. The lower limit $\sigma_{\rm min}$ is the maximum value between the local pressure scale height $H$ and the Gaussian dust width. To estimate the pressure scaleheight, we use the temperature profile from Eq.~\ref{eq:temp}, and derive $H$ knowing that:

\begin{equation}
    H(R) = \sqrt{ \frac{k_{\rm B}T}{\mu m_{\rm H}} \frac{R^3}{GM_*} },
\end{equation}
with $\mu$ being the molecular weight and $m_{\rm H}$ being the hydrogen mass. The upper limit $\sigma_{\rm max}$ is set such that the distance between the two rings is larger than the full-width-half-maximum of the gas Gaussian profile. Given that the rings in J1610 are likely unresolved even in the $(u,v)$-plane analysis (see Sec.~\ref{sec:galario}), we can determine upper limits only on $\alpha/{\rm St}$. In particular, we obtain an upper limit of $\alpha/{\rm St}_{\rm max}\gtrsim15.3$ for B41 in J1610, and a lower limit of $\alpha/{\rm St}_{\rm min}\lesssim0.29$ for B69 in LkCa 15. These values are in general agreement with the results on the DSHARP sample \citep{dullemond2018}, possibly suggesting that dust confinement within rings in full or transition disks may be similar. Additional information could be derived by having higher angular resolution observations on J1610 to radially resolve the rings, having multi-wavelengths datasets to translate ${\rm St}$ into a physical grain size, and having direct measurements of the ring widths in the gas component determining the pressure profile through kinematical analysis \citep[e.g.,][]{teague2018,rosotti2020}.

\paragraph{Planet-disk interactions.} In Sec.~\ref{sec:hydro} we explored the possibility that multiple rings in transition disks are footprints of an embedded planet, while we have also shown that disk thermodynamics alone can drastically change the morphology in the mm-continuum emission. What is the cooling timescale in disks?  Adopting the gray atmosphere approximation \citep{hubeny1990} and assuming that the disk temperature is dominated by stellar irradiation, the cooling timescale can be estimated as \citep[see e.g.,][]{zhu2015}
\begin{equation}
        t_{\rm cool}  =   {{3 \Sigma_{\rm g} c_V} \over {64 \sigma_{\rm SB} T_{\rm mid}^3}} {1+ \tau_R^2 \over \tau_R}
\end{equation}
or 
\begin{eqnarray}
        \beta & = & t_{\rm cool} \Omega \\
        & = & 0.024 \left( {\Sigma_{\rm g} \over 35~{\rm g~cm}^{-2}} \right)  \left( {T_{\rm mid} \over 34~{\rm K}} \right)^{-3} \left( M_* \over 1.25~M_\odot \right)^{0.5} \left( R \over 70~{\rm au} \right)^{-1.5}\\
        & & \times  \left( 1+\tau_R^2 \over \tau_R \right),
\end{eqnarray}
where $c_V \equiv (\gamma-1)\mathcal{R}/\mu$ is the specific heat capacity of gas, $\mathcal{R}$ is the gas constant, $\mu=2.4$ is the mean molecular weight, and $\tau_R = \kappa_R \Sigma_{\rm g}/2$. With the initial disk conditions described in Sec.~\ref{sec:hydro} and a range of Rosseland mean opacity value of $\kappa_R = 0.1 - 10~{\rm cm}^2~{\rm g}^{-1}$, the cooling parameter $\beta$ at 70~au in the LkCa~15 disk is $\sim0.07 - 4.3$. Note that we adopted a range of Rosseland mean opacity value as it requires a proper knowledge of the dust size distribution which depends upon a variety of physical processes and parameters, including the dust fragmentation velocity, the level of disk turbulence, dust internal density, and underlying gas density and temperature \citep{birnstiel_18}. In addition, the dust-to-gas mass ratio as well as the dust size distribution can evolve over time as a planet redistributes dust particles around its orbit. Nevertheless, this simple estimation shows that deviations from the isothermality, which previous studies have shown to occur for $\beta \gtrsim 0.01$ \citep{miranda2019,zhang2019}, are likely to happen in protoplanetary disks and should be considered when modelling their observations with hydrodynamical simulations.

In addition to disk's thermodynamic properties, many other parameters and physical processes also play a role in determining the location and appearance of the mm-rings generated by the planet-disk interactions. Among others, the underlying gas density structure, disk temperature, turbulence, planet mass, grain properties, dust feedback, orbital migration. This shows how challenging it is to retrieve planet mass and location by analysing the properties of rings in disks (in the hypothesis that these are the outcome of planet-disk interactions). At the same time, the resemblance of the morphological features observed in disks with the results of hydrodynamical simulations of planets-disk interactions is surprising, and suggests that rings in disks are (at least in part) generated by planets. The level and morphology of substructures in transition disks, and in particular LkCa~15 and J1610, suggest that the observed morphology can be caused by low mass planets that are formed within massive dust rings. Even though we have been agnostic on the mechanism generating the inner cavity by using a prescribed gas surface density, in the assumption of massive planets clearing the inner cavity it is reasonable to speculate that the resulting dust trapping can generate secondary planet formation from the inside-out, with planetary cores forming in dust traps created by the inner and more massive planet. However, we point out that planetary core formation via pebble accretion within dust rings at large distances from the central star ($\gg10$\,au) may be challenging due to the long dynamical timescales \citep[e.g.,][]{morbidelli2020}.

\paragraph{Snow lines.} While our model shows that a low mass planet is able to reproduce the observed features, planet-disk interactions are not the only means to form annular structure in disks. Among the numerous models proposed, dust properties and opacities are expected to vary at the ice lines of abundant chemical species, with physical mechanisms such as outward diffusion and sintering accentuating the contrast outside the ice-lines \citep{stevenson1988,sintering}. While surveys of disk annular substructures did not find any correlation between the location of the rings with estimated ice lines location in disks when using simple temperature prescriptions \citep{huang2018b,long2018}, indicating a different origin for most rings, some of them might still be induced at condensation fronts. N$_2$H$^{+}$ is a molecule known to be abundant between the CO and N$_2$ ice lines \citep[e.g.,][]{qi2013,vanthoff2017}, and most importantly it traces the location of the ice lines without needing to assume a temperature or a desorption temperature of the two molecules. \citet{qi2019} observed N$_2$H$^{+}$ in LkCa 15, showing that this ion is particularly abundant between $58^{+6}_{-10}$ and $88^{+6}_{-4}\,$au. These radii are just inwards of the B69 and B100 rings, respectively, and it may suggest that the two rings are formed by condensation of these two abundant molecules. However, as noted by \citet{huang2020} in the similar case of GM Aur, the scattered light data by \citet{thalmann2016} do not show clear rings co-located with the ones observed in ALMA. This seems to favour a planetary origin of the rings, rather than chemical, since in the first case rings in scattered light are predicted to be less pronounced than at mm wavelengths, whereas in the second case the opposite is expected \citep{pinilla2017}. Interestingly, both GM Aur and LkCa~15 show a faint outer ring in N$_2$H$^+$ emission at 200 and 220\,au, respectively \citep{qi2019}, and at the same time show an outer shallow ring/shoulder in the continuum intensity profile  interior of the molecular ring (at $\sim170$ and $\sim175\,$au, respectively), as shown in this paper and in \citet{huang2020}. It is likely that the outer N$_2$H$^+$ ring is related to a radial thermal inversion in the proximity of the dust outer radius which in turns leads to CO being released into gas phase again \citep[e.g.,][]{facchini2017}; however, it would be worth exploring whether there is any potential connection between outer CO snow-lines and shallow rings close to edge of the mm-emission, analogously to what is expected to happen at the inner CO-snowline but in the opposite radial direction. No observations of molecular species rather than CO are available for J1610, thus estimates of ice lines locations would still suffer from the large uncertainties attributed to disk temperature retrieval.

It is worth mentioning that there can be alternative scenarios to the planet-disk interaction and snow lines cases. For example, \citet{pinilla2016} modelled the grain size evolution in pressure maxima generated by a dead-zone \citep[i.e. a radial discontinuity in the viscosity parameter governing radial advection of gas parcels in a disk][]{gammie2006,regaly2012,flock2015}, showing that substructure in the dust properties at the dead zone edge may occur on Myr timescales (see their Fig. 4). The present observations cannot distinguish between these different models.

\paragraph{Inner disk and azimuthal asymmetries.}
Both LkCa~15 and J610 exhibit an inner disk via an IR excess \citep[e.g.,][]{espaillat2008,ansdell2016}, 
and LkCa~15 shows detectable mm continuum emission at the center of the cavity in the ALMA image. In the assumption that this is due to dust in the proximity of the star, by using the same dust opacity as above and a temperature of $100\,$K, the inner disk shows a dust mass of $\sim0.4$ Moon masses in the optically thin approximation. Since the emission seems unresolved at our spatial resolution, the flux density of the inner disk can be related to an inner disk size in the optically thick approximation, with a radius of $\sim 0.15\,$au for a temperature of $100\,$K. By using a gas-to-dust ratio of 100, the estimated mass in gas of the inner disk emitting at $1.3\,$mm is $\sim1.5\times10^{-6}\,M_\odot$, which would be entirely accreted in $\sim3000\,$yrs at the current accretion rate, in a too short timescale to be realistic. This can be easily overcome since the mass estimate we computed is a lower limit in the optically thin assumption, and in particular in case the inner disk is constantly replenished by the outer disk through material flowing into the cavity.

Interestingly, both Lkca~15 and J1610 are known to host variability in their optical light curve, and have been both classified as dippers \citep[e.g.,][]{ansdell2016,alencar2018}. Dippers are pre-main sequence stars that host quasi-periodic or episodic dimming in the lightcurve, with durations of about one day \citep[e.g.,][]{alencar2010,cody2014}. Different models have been invoked to explain the photometric behaviour, in general requiring small amount of dust close to the dust sublimation radius to be interposed along the line of sight generating partial occultation of the central star. Among many, proposed models are magnetospheric accretion lifting dust along the accretion column \citep{bouvier1999}, dust laden winds, thermal instabilities and/or misaligned inner disks. Many of these mechanisms rely on the inner disk to be close to edge-on \citep[e.g.,][]{bodman2017,cody2018}, in order for the mechanism to be able to displace dust along the line of sight. However, \citet{ansdell2020} have shown that the distribution of outer disk ($R\gtrsim10\,$au) inclinations of dippers as retrieved by ALMA imaging is isotropic, suggesting that outer disk inclination is not a relevant parameter to determine the 'dipper' status of young stars. 
In both disks analysed in this paper, the residuals shown in Fig.~\ref{fig:map_mod_res_lkca15} show asymmetric features. Such features could be due both to thermal effects (and thus shadowing) or to local surface density effects, for example generated by decaying vortexes in low turbulence environments. LkCa~15 shows a North-West region that is brighter than the South-East, in agreement with the scattered light image and model by \citet{thalmann2015}. At the same time, J1610 presents evident negative residuals at PA of $\sim55^\circ$ and $\sim190^\circ$, and strong positive residuals in the South-East side of the disk. This feature is strikingly similar to what is observed in scattered light images of disks interpreted as hosting a misaligned inner disk \citep[e.g.,][]{marino2015,benisty2018,muto-arena2019} with a misalignment of $\lesssim30^\circ$ \citep{facchini2018_warp}. Dedicated radiative transfer models and analysis of the gas kinematics in the inner regions of the central cavity will be key in confirming and characterizing the presence of misaligned inner disks in both systems.

\section{Conclusions}
\label{sec:conclusions}
In this paper, we present and analyze high angular resolution ALMA continuum observations of two transition disks, namely LkCa~15 and 2MASS~J16100501-2132318. Our main results are summarized as follows:
\begin{itemize}
    \item Both disks are structured and the broad rings previously imaged at lower  resolution resolve in multiple rings. LkCa~15 has rings located at $\sim$47, 69 and 100\,au, with widths $\sim$9, 6, 14\,au, respectively, and possibly another faint ring located further out at $\sim175\,$au. J1610 shows two rings at $\sim$29 and 41 au, with widths of $\sim$2, 3\,au at the current resolution. Only the two outer rings in LkCa~15 are radially resolved at the present angular resolution.
    Both disks also host an extended component covering a broad range of radii. 
    \item Rings appear to be marginally optically thick assuming negligible albedo of the dust grains, with optical depths at peaks ranging between $\sim$0.3 and 0.6. All rings are more massive than 10\,$M_\oplus$, and can be as massive as $\gtrsim150\,M_\oplus$. The properties of the rings are similar to the ones observed in  protoplanetary disks with no cavity (i.e. not transition) when imaged at high angular resolution.
    \item While evidence for the presence of an inner disk has already been inferred from the near-IR excess in the SED \citep[e.g.,][]{espaillat2008}, only LkCa~15 shows evidence for a compact millimeter emission inside the cavity indicating a dust-laden inner disk. At our sensitivity, we do not detect such inner emission in J1610.
    \item Both disks show azimuthal asymmetries in the brightness temperature distribution, and they resemble theoretical predictions of shadowing from misaligned inner disks. Increasing the sample of transition disks hosting dipper stars may help elucidate the link between inner and outer disks in objects with dynamically variable inner regions.
    \item Comparison with hydrodynamical simulations show that a planet embedded in the outer disk can lead to the formation of multiple rings beyond a cavity. However, we find that the thermodynamics can dramatically affect the number and appearance of these rings, suggesting that planet and disk properties inferred from models should be interpreted with caution. The uncertainty of the disk cooling adds to other uncertainties from planet migration, planet formation timescales, disk turbulence, among others that have been previously studied.
    \item The radii of the two prominent rings in LkCa~15 are in broad agreement with observations locating the CO and N$_2$ snowlines from N$_2$H$^+$ observations \citep{qi2019}, suggesting that chemistry may be causing (or facilitating) the intensity contrast observed in the radial profile.
    \item The rings in both disks appear as favourable locations where streaming instability can occur, leading to planetesimal (and possibly subsequent planetary core) formation. The qualitative agreement between the hydrodynamical simulations and the observed morphology suggests that massive rings in transition disks may be the best environments to characterize the conditions of formation of planetary cores. If the inner cavity is cleared by a yet unseen massive planet, we may be witnessing a second-generation planet formation event triggered by the dust trap generated by the inner body.
\end{itemize}

Future observations at even higher angular resolution may pin down the nature of the inner shoulder (B49) observed in LkCa~15. At the same time, high angular resolution observations can provide better constraints on the efficiency of dust trapping and on the dust diffusion within rings. Multi-frequency high resolution observations and gas kinematics studies will be particular beneficial to shed light on the dust trapping mechanisms in transition disks.


\begin{acknowledgements}
We are thankful to Marco Tazzari, Leonardo Testi and Antonella Natta for fruitful discussions. This paper makes use of the following ALMA data: ADS/JAO.ALMA\#2018.1.01255.S. ALMA is a partnership of ESO (representing its member states), NSF (USA), and NINS (Japan), together with NRC (Canada),  NSC and ASIAA (Taiwan), and KASI (Republic of Korea), in cooperation with the Republic of Chile. The Joint ALMA Observatory is operated by ESO, AUI/NRAO, and NAOJ.  S.F. acknowledges an ESO Fellowship, and is grateful to JAO for hosting a month long visit which was greatly beneficial for completing the paper. M.B. acknowledges funding from ANR of France under contract number ANR-16-CE31-0013 (Planet Forming Disks). J.B. acknowledges support by NASA through the NASA Hubble Fellowship grant \#HST-HF2-51427.001-A awarded  by  the  Space  Telescope  Science  Institute,  which  is  operated  by  the  Association  of  Universities  for  Research  in  Astronomy, Incorporated, under NASA contract NAS5-26555, and computing resources provided by the NASA High-End Computing Program through the NASA Advanced Supercomputing Division at Ames Research Center. P.P. acknowledges support provided by the Alexander von Humboldt Foundation in the framework of the Sofja Kovalevskaja Award endowed by the Federal Ministry of Education and Research. AWM was partially supported through NASA's Astrophysics Data Analysis Program (80NSSC19K0583). This project has received funding from the European Union's Horizon 2020 research and innovation programme under the Marie Sklodowska-Curie grant agreement No 823823 (Dustbusters RISE project), and was partly supported by the Deutsche Forschungs-gemeinschaft (DFG, German Research Foundation) - Ref no. FOR 2634/1 TE 1024/1-1.
\end{acknowledgements}


\bibliographystyle{aa}
\bibliography{references}

\begin{appendix}

\section{Stellar properties of J1610}
\label{sec:app_sed}
\begin{figure}[tb]
    \centering
    \includegraphics[width=0.49\textwidth]{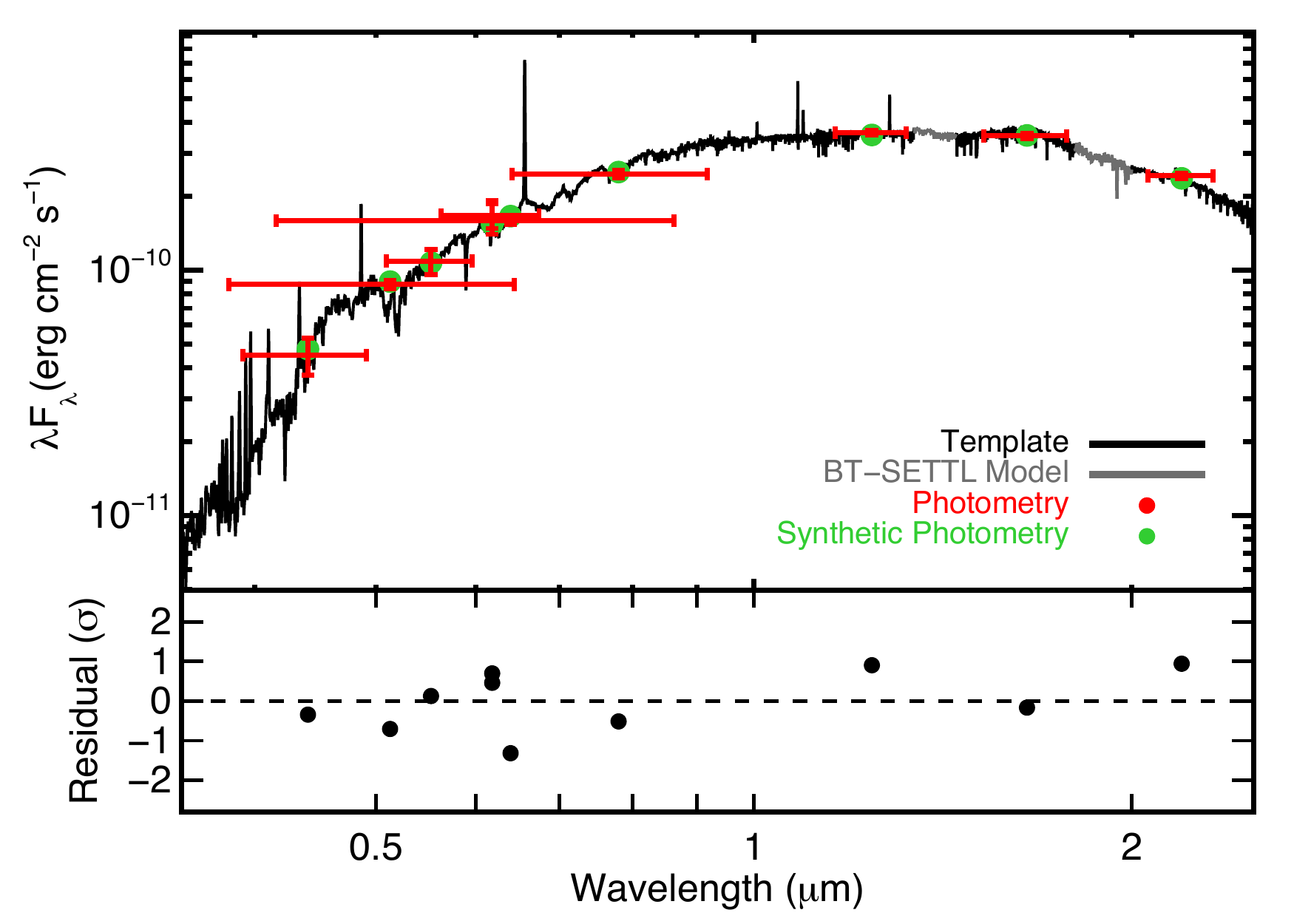}
    \caption{Best-fit spectral-energy distribution for J1610 created by generating synthetic photometry (green) from our observed spectra (black) and model atmosphere (grey) and locking it to the literature photometry (red). Vertical error bars on the observed photometry correspond to uncertainties (include an estimate of the stellar variability) and horizontal errors represent the width of the filter profile. The bottom panel shows the residual photometry in units of standard deviations. We de-redden the calibrated spectrum before computing the bolometric flux, but the spectrum and photometry shown here are still reddened. }
    \label{fig:sed}
\end{figure}

To determine the luminosity of J1610 we fit the spectral-energy distribution following the methodology from \citet{Mann2016b}, which we briefly summarize here. We use the spectra taken in \citet{ansdell2016}, which we absolutely calibrate using literature photometry from 2MASS \citep{Skrutskie2006}, Gaia DR2 \citep{Evans2018}, APASS \citep{Henden2012}, and CMC15 \citep{CMC15}. We simultaneously fit for reddening by comparing the spectrum to un-reddened templates of stars in nearby young moving groups observed with the same instruments and setup. We fill in gaps in the observed spectra (e.g., beyond 2.2 microns) by interpolating over a grid of BT-SETTL model atmospheres \citep{Allard2013}. We calculate the bolometric flux from the integrated and un-reddened spectrum, and the luminosity from the bolometric flux and Gaia DR2 distance. We show the best-fit reddened spectrum in Figure~\ref{fig:sed}. Final uncertainties account for errors in the flux calibration of the spectra, variability in the observed photometry, and differences in reddening from different selections of the un-reddened template. Our final value is $L_*=0.46\pm0.03L_\odot$, with $E(B-V)=0.26\pm0.06$, and $T_{\rm{eff}}=3950\pm80$\,K.

\section{Radial profiles and visibility modelling}
\begin{figure*}
\begin{center}
\includegraphics[width=0.48\textwidth]{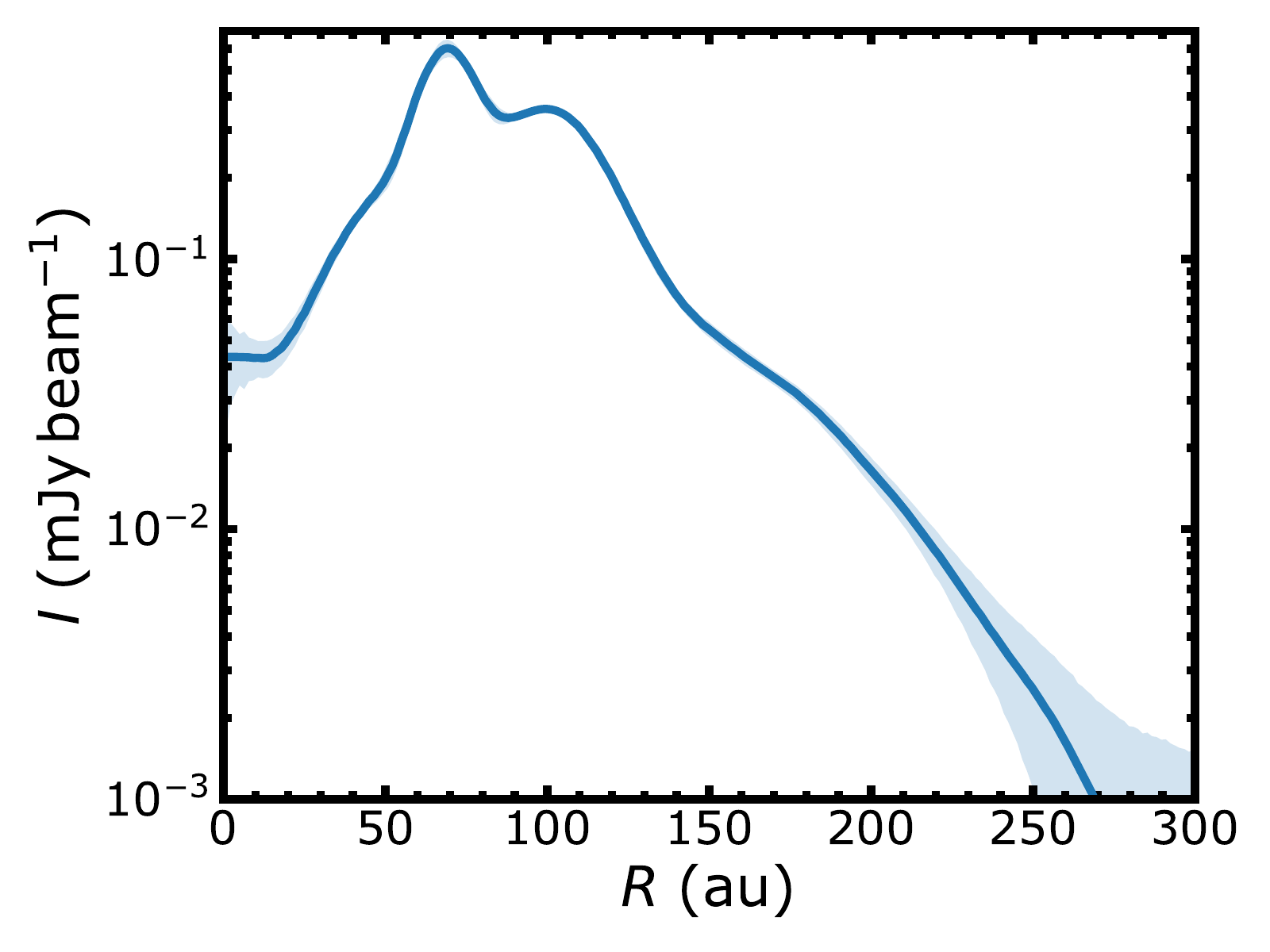}
\includegraphics[width=0.48\textwidth]{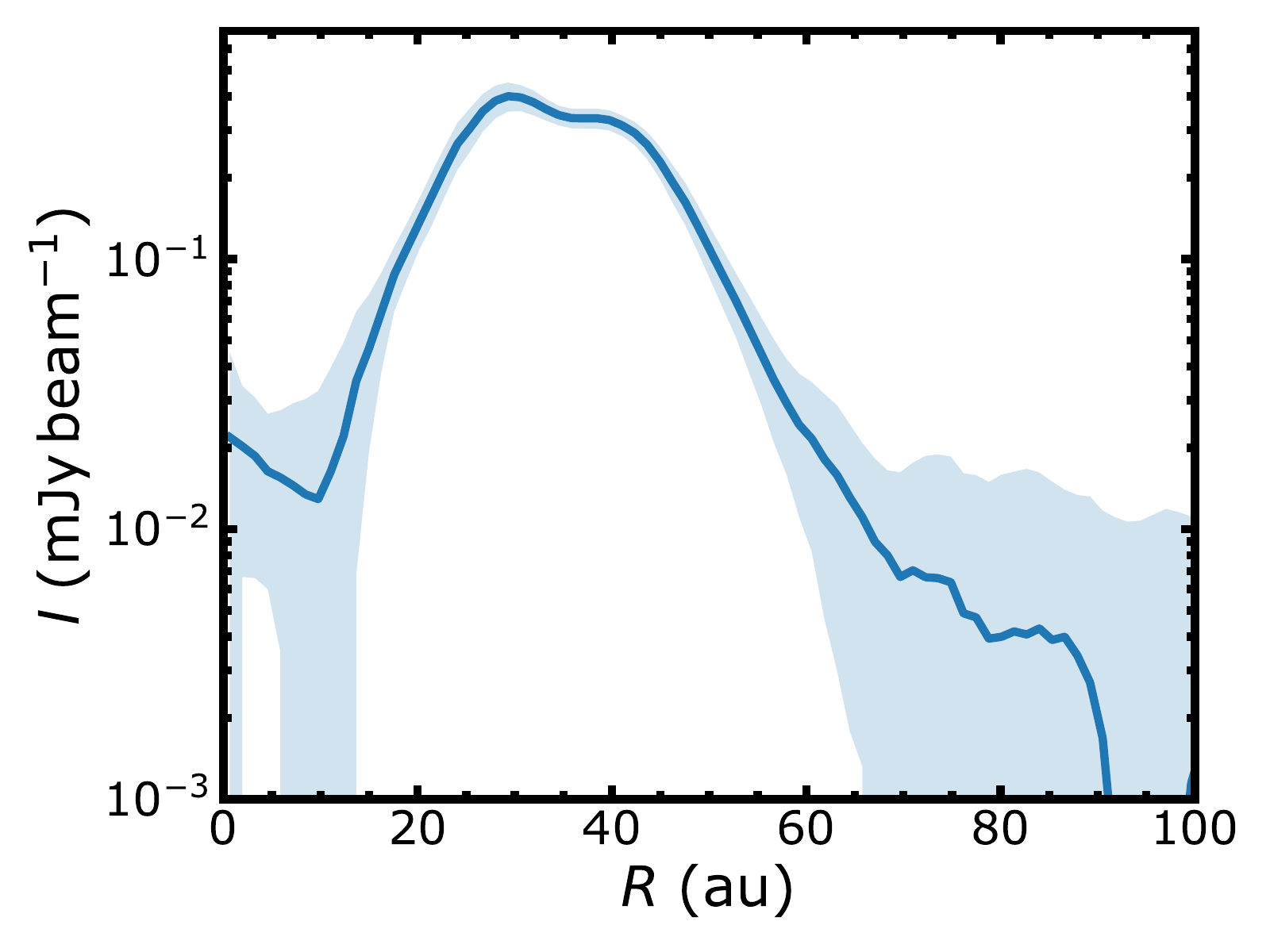}
\end{center}
\caption{Deprojected and azimuthally averaged radial intensity profile of the continuum emission of LkCa~15 (left panel) and J1610 (right panel) as in Fig.~\ref{fig:rad_prof}. The blue ribbon shows the quadratic sum of the standard deviation of the intensity across pixels in each radial bin and the rms of the observations divided by the square root of the number of independent beams sampling the same radial bin in the azimuthal direction.}
\label{fig:rad_prof_log}
\end{figure*}

\begin{figure*}
\begin{center}
\includegraphics[width=0.48\textwidth]{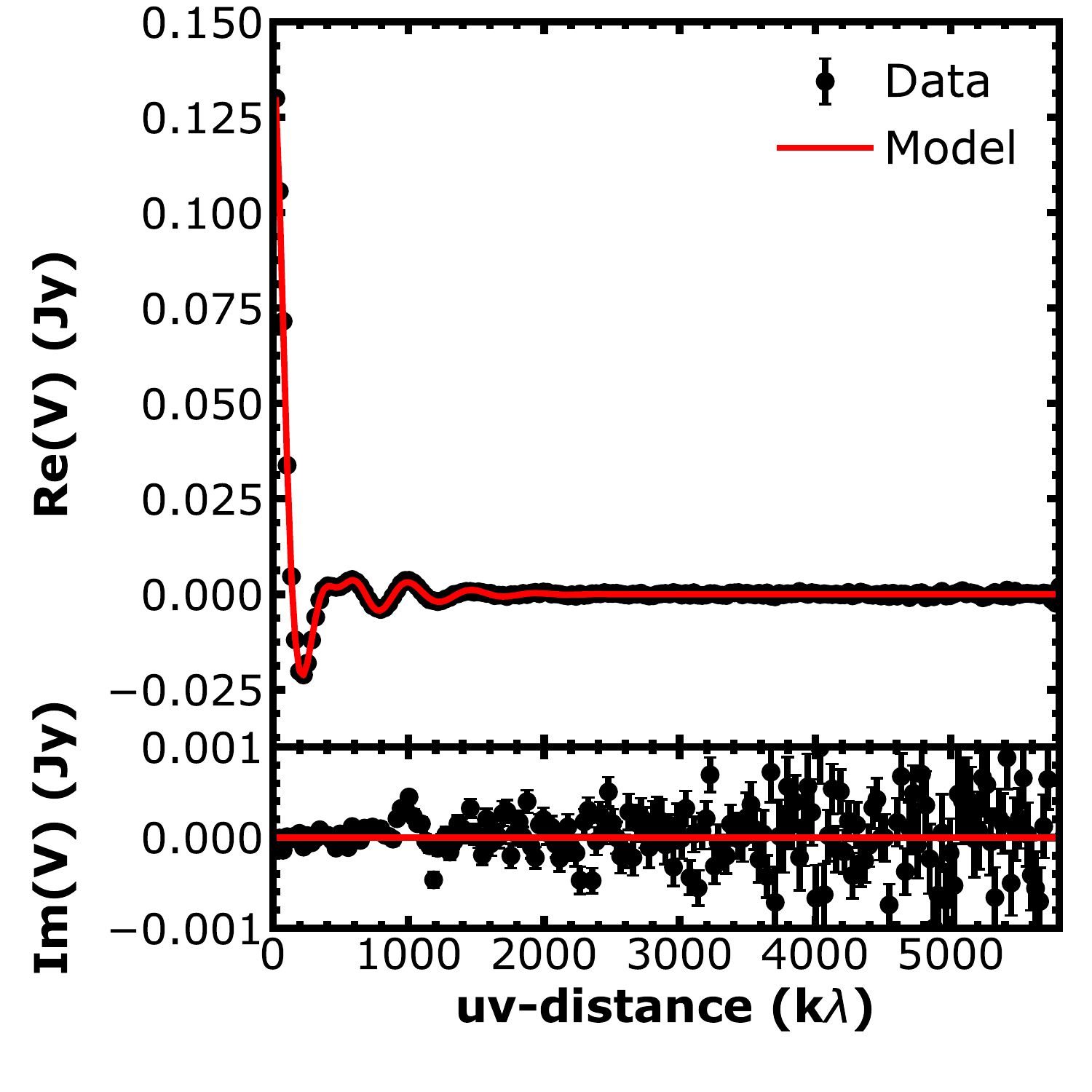}
\includegraphics[width=0.48\textwidth]{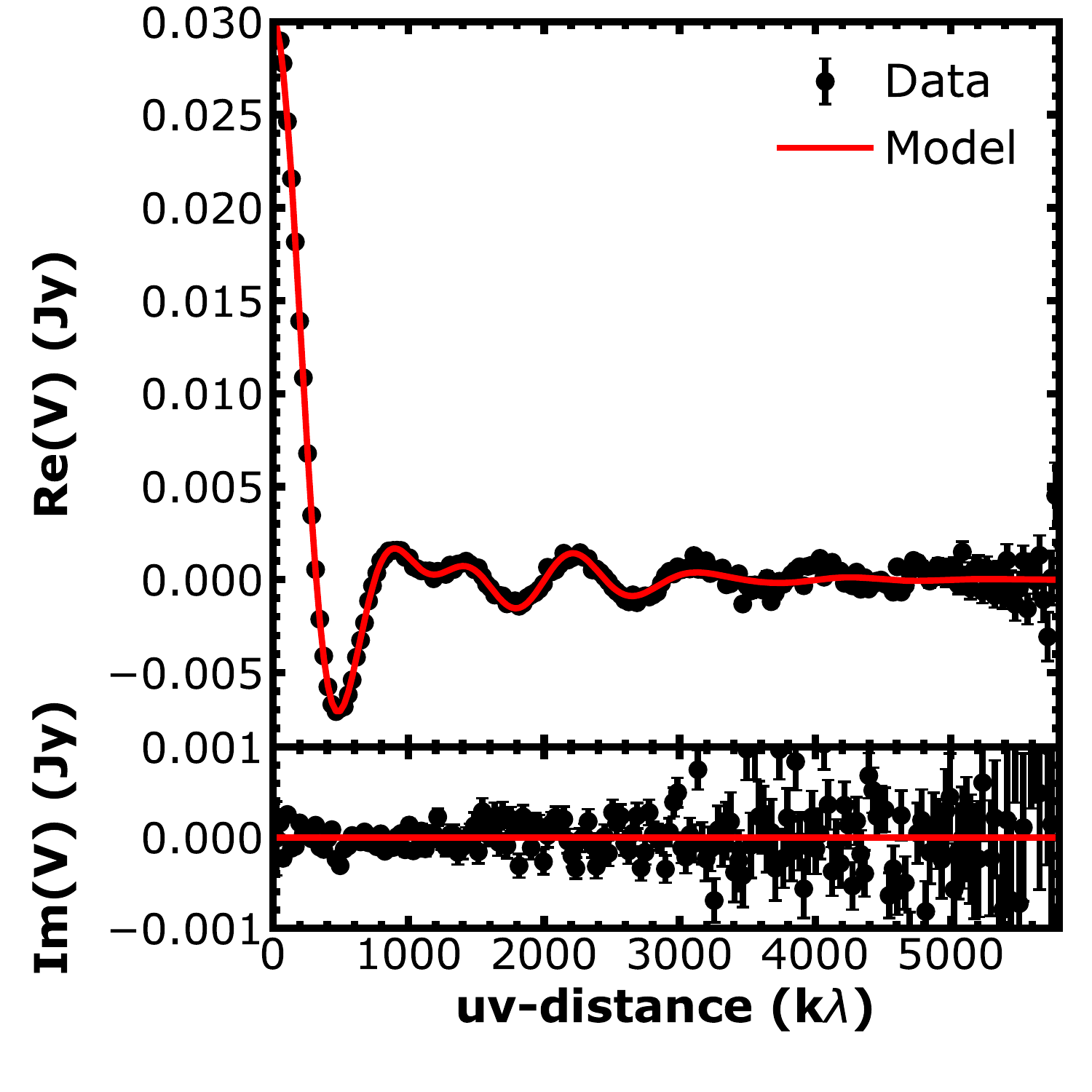}
\end{center}
\caption{Re-centered and deprojected visibililities of the data and best fit model for LkCa~15 (left panel) and J1610 (right panel). Error bars show $1\sigma$ uncertainties. The parameters of the best fits are listed in Tables~\ref{tab:LkCa15}-\ref{tab:J1610}. The plots have been made with the uvplot package \citep{uvplot}.}
\label{fig:visibilities}
\end{figure*}

Fig.~\ref{fig:rad_prof_log} shows the azimuthally averaged intensity profiles of LkCa~15 and J1610 in log scale, to highlight the profile at large radii. Fig~\ref{fig:visibilities} shows the re-centered and deprojected visibilities of the data with the best fit model.

\section{Initial disk temperature profile from MCRT iterations}

\begin{figure*}
\begin{center}
\includegraphics[width=0.48\textwidth]{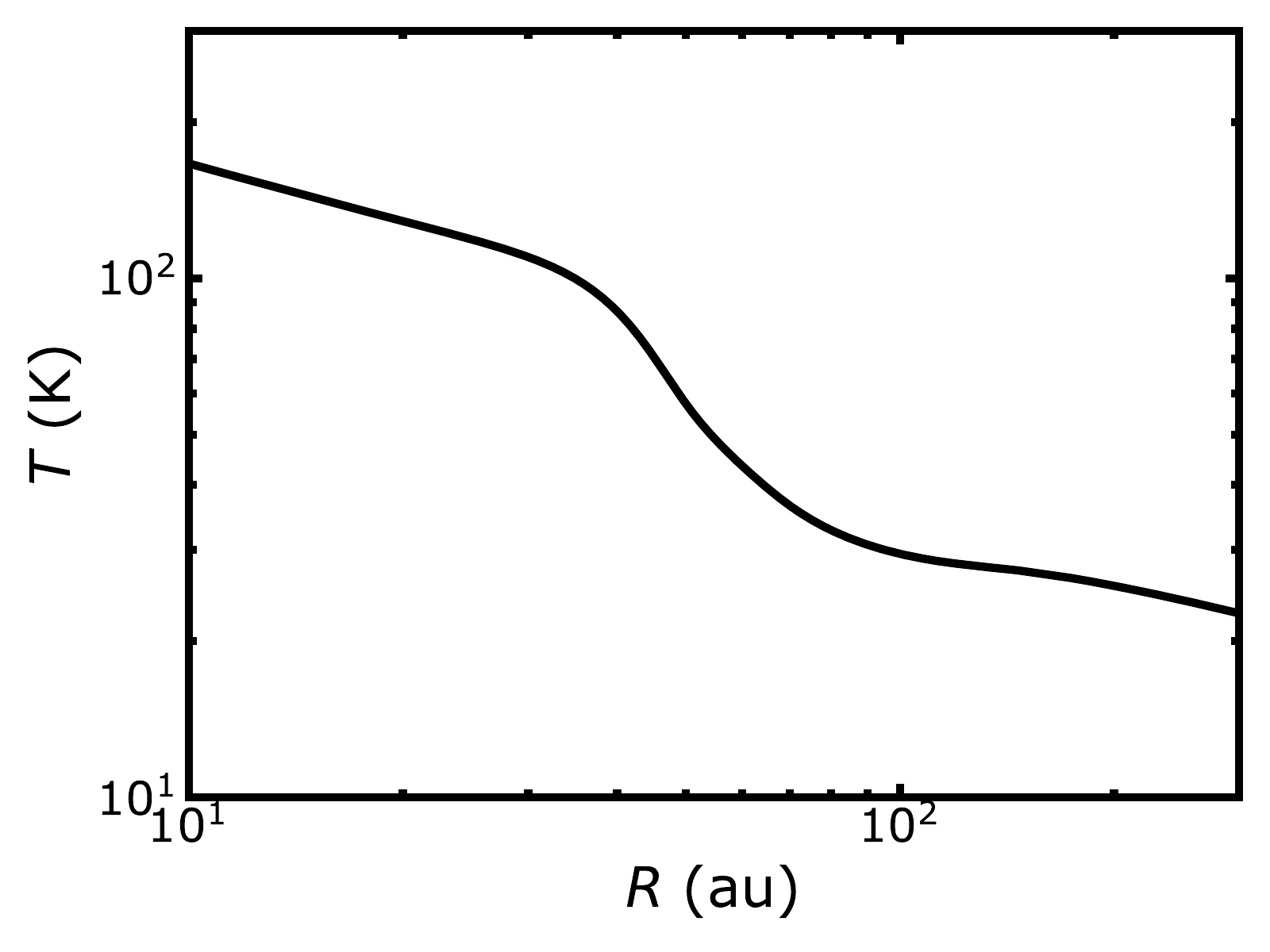}
\includegraphics[width=0.48\textwidth]{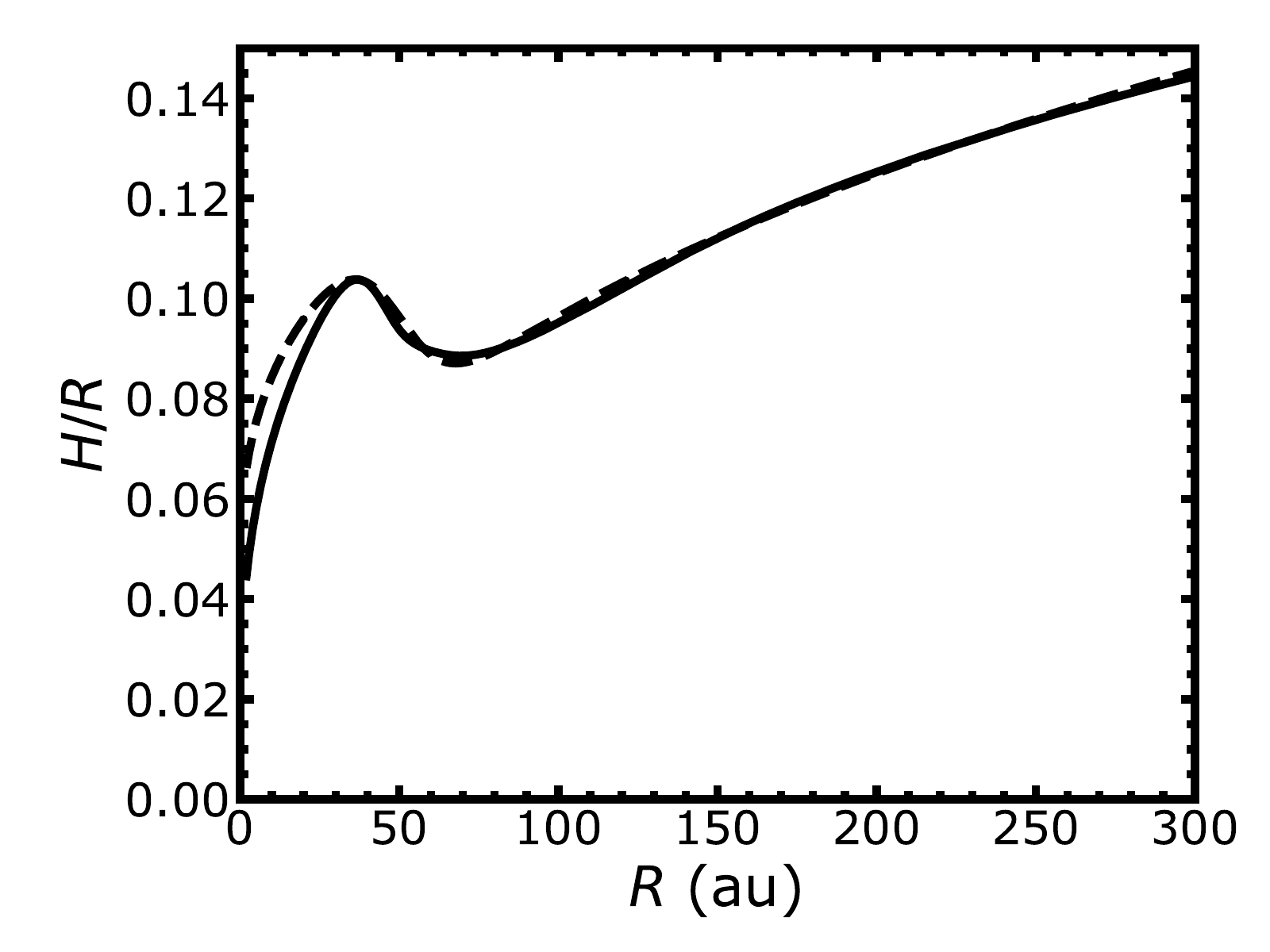}
\end{center}
\caption{Left: density-weighted, vertically integrated disk temperature profile from the MCRT iterations described in Sec. \ref{sec:set-up}. Right: disk aspect ratio $H/R$ corresponding to the temperature profile in the left panel, adopting a stellar mass of $1.25~M_\odot$ and a mean molecular weight of 2.4. The dashed curve shows a fit to the profile between 30 and 300~au, as expressed by Eq. \ref{eq:temperature}.}
\label{fig:disk_temperature}
\end{figure*}

Fig. \ref{fig:disk_temperature} shows the disk temperature profile obtained from the MCRT iterations described in Sec. \ref{sec:set-up} and the corresponding disk aspect ratio profile.

\section{On the point source within the LkCa~15 cavity}

\begin{figure*}
\begin{center}
\includegraphics[width=0.42\textwidth]{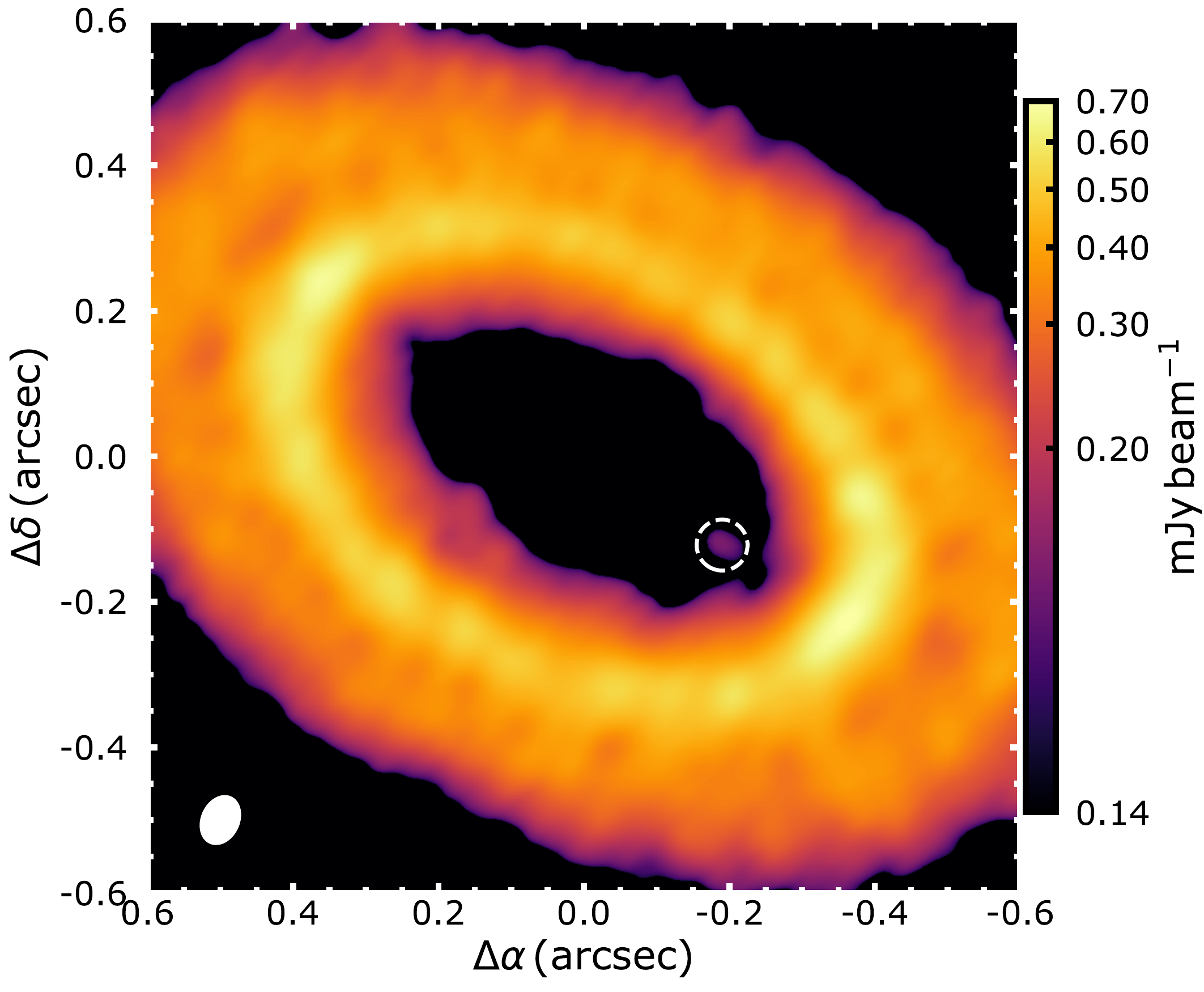}
\includegraphics[width=0.48\textwidth]{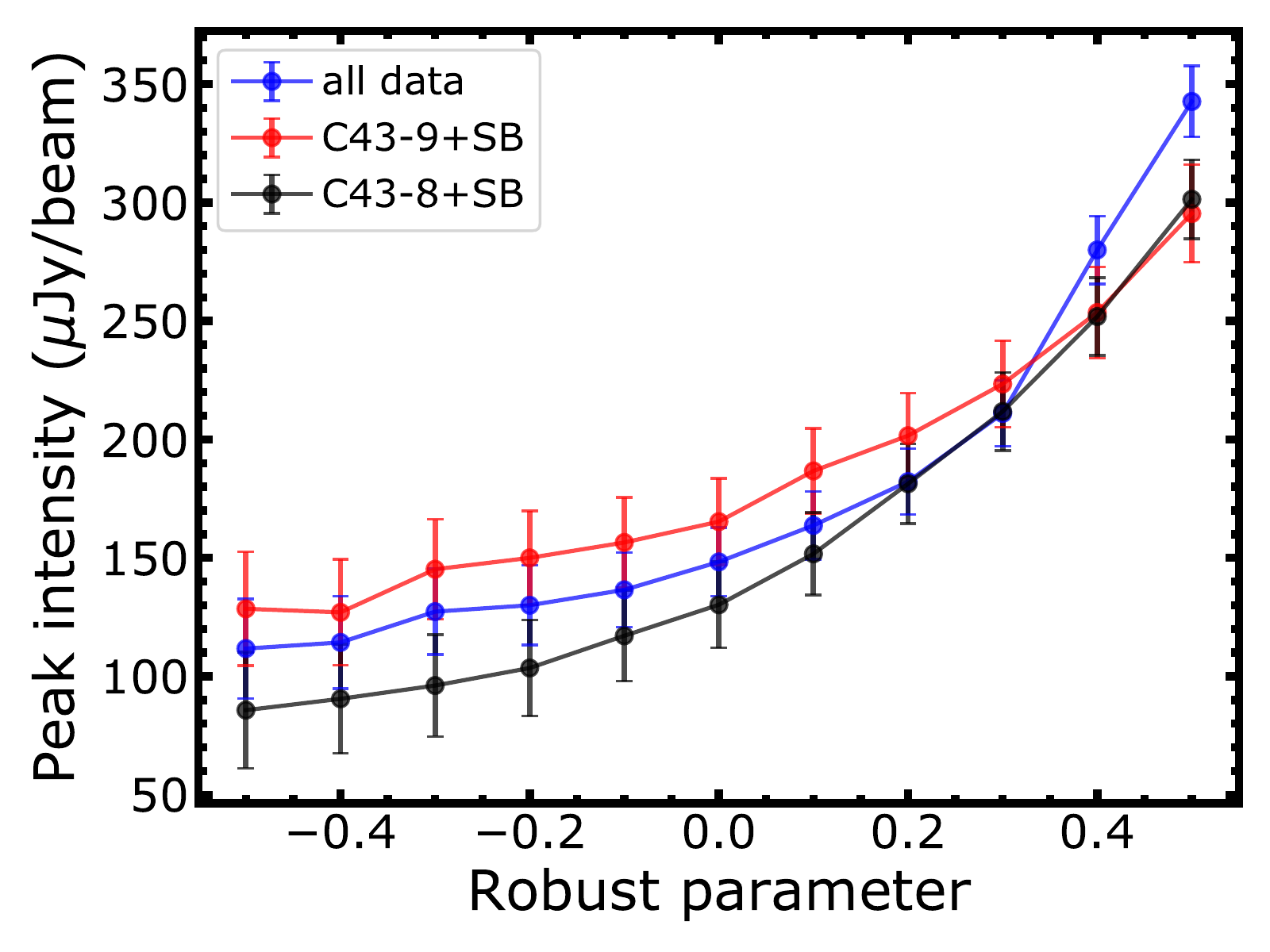}
\end{center}
\caption{Left: zoom onto the cavity of LkCa~15, with stretched color scale to emphasize the fainter emission within the cavity. The white dashed circle indicates the presence of a potential point source within the cavity. Right: peak intensity of the potential point source in images generated with Briggs weighting but different robust parameters. The angular resolution is decreasing going from the left to the right of the plot. Different colors show the same analysis performed on the full data set, or on datasets including the short baselines and one of the two long baselines execution blocks, with C43-9 being the execution block of July 13th, and C43-8 being the execution block of July 19th (labelled following the nominal configurations of the observations.)}
\label{fig:blob_lkca15}
\end{figure*}

Fig.~\ref{fig:blob_lkca15}, left, shows an image of LkCa~15 with color code stretched to highlight the point source like emission within the cavity. We imaged the disk by combining the short spacings observation with only one long baseline observations at the time. Interestingly, the point source like feature is clearly imaged when only the C43-9 observation is considered, but it is not apparent in the C43-8 plus short spacing image. This supports the discussion presented in Sec.~\ref{sec:images}, i.e. that the feature is an image artifact driven in particular by the C43-9 execution block. The right panel of Fig.~\ref{fig:blob_lkca15} shows the peak intensity within the area highlighted in the left panel, and how it varies with imaging parameters.

\end{appendix}

\end{document}